\documentclass[aps,prd,amsmath,amsfonts,amssymb,eqsecnum,nofootinbib,
  floatfix,secnumarabic,preprintnumbers,superscriptaddress,nobalancelastpage,
  onecolumn,notitlepage]{revtex4}
\pdfoutput=1
\usepackage{subfigure,color,graphicx,hyperref,dsfont,slashed,multirow}
\hypersetup{bookmarks=true,unicode=true,pdftoolbar=true,pdfmenubar=true,
  pdffitwindow=false,pdfstartview={FitH},pdftitle={VLQMSSM},
  pdfauthor={J.~Araz,~S.~Banerjee,~M.~Frank,~A.~Goudelis,~B.~Fuks},
  pdfsubject={VLQ},
  pdfcreator={Creator},pdfproducer={Producer}, pdfkeywords={Non-minimal 
  supersymmetry}{Beyond the Standard Model Physics}{sneutrino dark matter},
  pdfnewwindow=true,colorlinks=true,
  linkcolor=blue,citecolor=magenta,filecolor=magenta,urlcolor=cyan}
\newcommand{\be}{\begin{equation}}
\newcommand{\ee}{\end{equation}}
\def\bsp#1\esp{\begin{split}#1\end{split}}
\def\bpm{\begin{pmatrix}} 
\def\epm{\end{pmatrix}} 

\begin{document}
\date{\today}
\preprint{CUMQ/HEP 199, IPPP/18/95}

\title{Dark matter and collider signals in an  MSSM extension with vector-like multiplets}
\author{Jack Y. Araz}
\email{jack.araz@concordia.ca}
\affiliation{Concordia University 7141 Sherbrooke St. West, Montreal, QC,
 CANADA H4B 1R6}

\author{Shankha Banerjee}
\email{shankha.banerjee@durham.ac.uk}
\affiliation{Institute for Particle Physics Phenomenology, Department of Physics, Durham University, Durham
DH1 3LE, United Kingdom}

\author{Mariana Frank}
\email{mariana.frank@concordia.ca}
\affiliation{Concordia University 7141 Sherbrooke St. West, Montreal, QC,
 CANADA H4B 1R6}

\author{Benjamin Fuks}
\email{fuks@lpthe.jussieu.fr}
\affiliation{Sorbonne Universit\'e, CNRS, Laboratoire de Physique Th\'eorique et Hautes \'Energies, LPTHE, F-75252 Paris, France}
\affiliation{Institut Universitaire de France, 103 boulevard Saint-Michel, 75005 Paris, France}

\author{Andreas Goudelis}
\email{andreas.goudelis@lpthe.jussieu.fr}
\affiliation{Sorbonne Universit\'e, CNRS, Laboratoire de Physique Th\'eorique et Hautes \'Energies, LPTHE, F-75252 Paris, France}
\affiliation{Sorbonne Universit\'e, Institut Lagrange de Paris (ILP), 75014 Paris, France}
\vspace{10pt}
\begin{abstract}
Motivated by grand unification considerations, we analyse a simple extension of
the minimal supersymmetric standard model with additional pairs of vector-like
chiral supermultiplets. We focus on the so-called LND setup, which enlarges the
particle content of the minimal model by two vector-like pairs of weak doublets
(one pair of leptons and one pair of down-type quarks) and one vector-like pair
of neutrino singlets. Imposing collider and low-energy constraints, sneutrinos
and neutralinos both emerge as possible lightest supersymmetric particles and
thus dark matter candidates. We perform a complete analysis of the dark
sector and study the viability of these neutralino and sneutrino dark matter
options. We show that cosmological considerations (the dark matter
relic abundance and its direct and indirect detection signals) restrict
neutralino dark matter to exhibit similar properties as in the minimal
supersymmetric standard model,
and impose the sneutrino dark matter candidate to be singlet-like, rather than
doublet-like. Allowing the mixing of the fermionic component of the
new supermultiplets with the Standard Model third generation fermions, we
moreover demonstrate the existence of collider signals that are distinguishable
from other, more minimal, supersymmetric scenarios by virtue of an enhanced
production of events enriched in tau leptons. We furthermore show that this
signature yields robust LHC signals, that could potentially be differentiated from the
background in future data.
\end{abstract}

\maketitle
\section{Introduction}\label{sec:intro}

During the past few decades, supersymmetry has gained the status of one of the best theoretically-motivated scenarios for physics beyond the Standard Model (SM). Under specific conditions it can address the hierarchy problem, achieve gauge coupling unification and, when $R$-parity is conserved, provide natural dark matter candidate(s). However, supersymmetric models of particle physics have been under assault both from collider searches and from direct and indirect dark matter detection experiments. The measured value of the Higgs-boson mass seems to require TeV-scale scalar quarks, a fact further strengthened by the results of direct sparticle searches at the LHC. The null results from dark matter searches have put substantial pressure on (light) neutralino dark matter and, since long ago, have wiped out left-handed sneutrinos as phenomenologically-viable dark matter candidates. Furthermore, in its minimal version, supersymmetry fails to explain neutrino masses. Before, however, abandoning low-scale supersymmetry, one may ask if some of these outstanding issues can be addressed in extensions of the minimal supersymmetric standard model (MSSM), while maintaining the attractive features of the latter and giving rise to novel signals at colliders and elsewhere.

A large variety of MSSM extensions have been studied in the past, including (but
not limited to) effective approaches~\cite{Casas:2003jx,Dine:2007xi}, as well as
minimal modifications of the MSSM particle content \cite{Arina:2007tm,%
Ellwanger:2009dp} or gauge group structure \cite{Batra:2003nj,Babu:2008ep}. In
Ref.~\cite{Martin:2009bg}, a less minimal approach was proposed, extending the
MSSM particle content by additional pairs of vector-like supermultiplets. The
advantage of this choice is that the Higgs-boson mass can be raised while
maintaining perturbative gauge coupling unification. The suggested models
involve either $\bf{5}+{\bf{\overline{5}}}$ complete representations of $SU(5)$
(the `LND' scenario) or $\bf{10}+{\bf{\overline{10}}}$ complete representations
of $SU(5)$ (the `QUE' scenario). Additionally, a `QDEE'
setup which does not contain complete multiplets of $SU(5)$ but still leads
to gauge coupling unification has also been envisaged. In this notation scheme,
the capital letters denote the nature of the extra supermultiplets relatively to
their MSSM counterparts carrying the same charge, colour and $B-L$ quantum
numbers.

The dark matter phenomenology of the QUE and QDEE models was already studied in
Refs.~\cite{Abdullah:2015zta,Abdullah:2016avr}. In both cases, the dark matter candidate is the lightest neutralino, much like in the MSSM, albeit with some interesting twists, and their phenomenology is rather similar. In this work we will focus on the third scenario, the so-called LND model. Although in this scenario the little hierarchy problem of the MSSM cannot be resolved \cite{Martin:2009bg}, the LND model presents some other attractive features.   As above-mentioned, it can lead to gauge coupling unification,
although this necessitates that the vector-like fermions have masses in the
600-1000~GeV window~\cite{Martin:2009bg}. The field content contains a pair of
vector-like neutrino singlets, whose fermionic components can be seen as a
sterile neutrino. This could consequently provide explanations for the hints of
neutrino oscillations at a higher frequency and for the differences between the
neutrinos and antineutrinos measured by the LSND and MiniBooNE
experiments~\cite{Athanassopoulos:1997pv, Aguilar-Arevalo:2013pmq}.
The model moreover features two potential dark matter candidates, the lightest
neutralino as well as the lightest singlet-like sneutrino. Furthermore, under a
specific configuration, it could give rise to additional contributions to the
anomalous magnetic moment of the muon, and large mixings between the new
fermions and the third generation SM fermions are allowed and can lead to
distinctive signals at the LHC.  With this as motivation, we expect the phenomenology of this model to differ from that of the QUE and QDEE models, and in this paper we perform an analysis of the dark matter constraints and collider implications for this model.

This paper is structured as follows. In Section \ref{sec:model} we present the
superfield content of the LND model, its superpotential and soft
supersymmetry-breaking Lagrangian, and detail the particle mixings that are
relevant for dark matter. In Section \ref{sec:scan} we describe the setup of our
parameter space exploration, and  provide information on the
experimental constraints that are imposed within our scan and the
computational tools that have been employed. In Section \ref{sec:DM} we study the dark matter phenomenology of our model, separately for the case of a neutralino and a sneutrino LSP. The consequences of the model at the LHC are explored in Section \ref{sec:detector}. Finally, we conclude in
Section~\ref{sec:conclusions}.

\section{Theoretical Framework} \label{sec:model}

\subsection{Field content and Lagrangian}
\label{sec:lagrangian}

The LND model is an extension of the
MSSM inspired by $SU(5)$ Grand Unification (GUT) considerations. We begin with the MSSM
chiral superfield content that contains three generations of quark ($Q$) and
lepton ($L$) weak doublets, as well as three generations of up-type quark ($\bar
U$), down-type quark ($\bar D$) and charged lepton ($\bar E$) weak singlets. In
our notations, the fermionic and scalar components of these supermultiplets read
\be\bsp
 & ~~~~~~~~~Q \equiv \big(q_L,\tilde q_L\big)
    \sim \Big(\mathbf{3}, \mathbf{2}, \frac16\Big) \ ,\hspace{3cm}
  L \equiv \big(\ell_L,\tilde \ell_L\big)
    \sim \Big(\mathbf{1}, \mathbf{2}, -\frac12\Big) \ ,\\
 & \bar U \equiv \big(u_R^c,\tilde u_R^\dag\big)
     \sim \Big(\bar{\mathbf{3}}, \mathbf{1}, -\frac23\Big) \ ,\qquad
  \bar D \equiv \big(d_R^c,\tilde d_R^\dag\big)
     \sim \Big(\bar{\mathbf{3}}, \mathbf{1},  \frac13\Big) \ ,\qquad
  \bar E \equiv \big(e_R^c,\tilde e_R^\dag\big) 
     \sim \Big(\mathbf{1}, \mathbf{1}, 1\Big) \ ,
\esp\label{eq:mssmmatter}\ee
where we also indicate their representation under the $G_{\rm MSSM}
\equiv SU(3)_c\times SU(2)_L \times U(1)_Y$ gauge group. The $^c$ superscript
indicates charge conjugation while the $_{L,R}$ subscripts refer to the
left- and right-handedness of the fermion. 

In the model considered in this work, the MSSM matter sector of Eq.~\eqref{eq:mssmmatter} is extended by
vector-like pairs of supermultiplets forming a complete $\mathbf{5}\oplus
\bar{\mathbf{5}}$ representation of $SU(5)$. Such a configuration allows to
keep a reasonable level of simplicity and to maintain perturbative gauge
coupling unification at high energy, with new states appearing at the TeV
scale~\cite{Moroi:1991mg,Moroi:1992zk,Babu:2004xg,Babu:2008ge}. Decomposing the
$\mathbf{5}\oplus \bar{\mathbf{5}}$ GUT supermultiplets in terms of the
$G_{\rm MSSM}$ gauge group, the chiral content of the model includes one pair of
vector-like leptons $(L_5, \bar L_5)$  in the fundamental representation of
$SU(2)_L$ and one pair of vector-like down-type quarks $(D_5, \bar D_5)$ 
in the trivial representation of $SU(2)_L$,
\be\bsp
  & L_5 \equiv \big(\ell_{5L},\tilde \ell_{5L}\big)
    \sim \Big(\mathbf{1}, \mathbf{2}, -\frac12\Big) \ , \qquad
  \bar L_5 \equiv \big(\ell_{5R}^c, \tilde{\ell}^\dag_{5R}\big)
     \sim \Big(\mathbf{1}, \mathbf{2},  \frac12\Big) \ , \\
  & D_5 \equiv \big(d_{5L},\tilde d_{5L}\big)
     \sim \Big({\mathbf{3}}, \mathbf{1},  - \frac13\Big) \ , \qquad
    \bar D_5 \equiv \big(d_{5R}^c, \tilde{d}_{5R}^\dag\big)
     \sim \Big(\bar{\mathbf{3}}, \mathbf{1}, \frac13\Big) \ .
\esp\ee
As in many GUT-inspired supersymmetric models, we also add a pair
of vector-like gauge singlets\footnote{This choice is consistent with minimal representations which do not violate lepton number.},
\be\bsp
  & N \equiv \big(N_L,\tilde N_L\big)
     \sim \Big(\mathbf{1}, \mathbf{1}, 0\Big) \ , \qquad
  \bar N \equiv \big(N_R^c, \tilde{N}^\dag_R\big)
     \sim \Big(\mathbf{1}, \mathbf{1}, 0\Big) \ ,
\esp\ee
that can be mapped to a pair of extra (s)neutrinos and find motivation in
dark matter and neutrino physics~\cite{Ellis:2010kf}. With
respect to the MSSM, the model features one extra down-type quark and two extra down-type squarks, 
one extra charged lepton and two additional charged sleptons
as well as two more neutrinos along with their accompanying four extra sneutrinos. The Higgs sector is identical to the MSSM one and consists of two weak doublets of Higgs supermultiplets
\be
    H_d \equiv \big(\tilde H_d, h_d \big)
    \sim \Big(\mathbf{1}, \mathbf{2}, -\frac12\Big) \ , \qquad
    H_u \equiv \big(\tilde H_u, h_u \big)
    \sim \Big(\mathbf{1}, \mathbf{2},  \frac12\Big) \ ,
\ee
that are sufficient to break $G_{\rm MSSM}$ down to  $U(1)_{\rm em}$ and
generate supersymmetric masses for all particles. Finally, the model includes three gauge
supermultiplets  (as the gauge group is the same as in the MSSM) that we denote
by
\be
  G\equiv \big(g, \lambda_{\tilde G}\big)
    \sim \Big(\mathbf{8}, \mathbf{1}, 0\Big) \ , \qquad
  W\equiv \big(w, \lambda_{\tilde W}\big)
    \sim \Big(\mathbf{1}, \mathbf{3}, 0\Big) \ , \qquad
  B\equiv \big(b, \lambda_{\tilde B}\big)
    \sim \Big(\mathbf{1}, \mathbf{1}, 0\Big) \ , \qquad
\ee
for the QCD ($G$), weak ($W$) and hypercharge ($B$) gauge groups.

The supersymmetry-conserving (non-gauge) interactions of the model are driven by
the superpotential $W_{\rm LND}$ that is written, assuming $R$-parity
conservation, as~\cite{Martin:2009bg}
\be\bsp \label{eq:superpotential}
  W_{\rm LND} = &\
     \mu\, H_u\cdot H_d + {\bf y_u}\, \bar U\, Q\cdot H_u
   - {\bf y_d}\, \bar D\, Q\cdot H_d - {\bf y_e}\, \bar E\, L\cdot H_d\\ &\
   + \mu_D\, D_5\, \bar D_5 + \mu_L\, L_5\cdot \bar L_5 + \mu_N\, N\, \bar N
   + k_N\, \bar N\, L_5\cdot H_u
   - h_N\, N\, \bar L_5\cdot H_d\\ &\
   - {\bf \varepsilon_D}\, \bar D_5\, Q\cdot H_d
   - {\bf \varepsilon_E}\, \bar E\, L_5\cdot H_d
   + {\bf \varepsilon_N}\, \bar N\, L\cdot H_u
   + {\bf \kappa_D}\, D_5\, \bar D
   + {\bf \kappa_L}\, L\cdot \bar L_5  \ ,
\esp \ee
where all flavour indices have been explicitly omitted for simplicity. The first
terms correspond to the MSSM superpotential in which $\mu$ denotes the MSSM
off-diagonal Higgs(ino) mass contribution and ${\bf y_u}$,  ${\bf y_d}$ and
${\bf y_e}$ stand for the up-type quark, down-type quark and charged lepton
Yukawa matrices in flavour space. Moreover, $\mu_D$, $\mu_L$ and $\mu_N$ are 
explicit masses for the non-MSSM fields and $h_N$ and $k_N$
stand for the Yukawa interactions of the non-coloured vector-like superfields.
The terms of the last line of the superpotential include new Yukawa couplings
driving the mixing of the Standard Model fermions with their vector-like
counterparts (the ${\bf \varepsilon_D}$, ${\bf \varepsilon_E}$ and
${\bf \varepsilon_N}$ vectors in flavour space) as well as direct mass mixing
terms (the ${\bf \kappa_D}$ and ${\bf \kappa_L}$ vectors in flavour space). While
these are strongly constrained by flavour data, such mixings have to be
non-vanishing to prevent the existence of unwanted cosmologically stable relics. For consistency with both flavour and cosmology constraints, in the
following we will assume the existence of a small mixing with the third
generation of Standard Model fermions only.

As in any realistic supersymmetric model, supersymmetry has to be softly
broken. The Lagrangian thus includes gaugino and scalar mass terms, as well as
bilinear and trilinear scalar interactions whose form is obtained from the
superpotential. The gaugino mass contributions read
\be\label{eq:softino}
  {\cal L}_{\rm soft}^{(\lambda)}= - \frac12 \Big(
    M_1 \lambda_{\tilde B} \cdot\lambda_{\tilde B} +
    M_2 \lambda_{\tilde W} \cdot\lambda_{\tilde W}+
    M_3 \lambda_{\tilde g} \cdot\lambda_{\tilde g}+{\rm h.c.}\Big) \ ,
\ee
where the $M_1$, $M_2$ and $M_3$ parameters represent the bino, wino and gluino
masses, and the scalar mass Lagrangian is given by
\be\bsp\label{eq:softscal}
  {\cal L}_{\rm soft}^{(\phi)} = &\ - 
    m_{H_d}^2h_d^\dag h_d - m_{H_u}^2 h_u^\dag h_u -
    {\bf m^2_{\tilde Q}}\, {\tilde q}_L^\dag\, \tilde q_L -
    {\bf m^2_{\tilde d}}\, {\tilde d}_R^\dag\, \tilde d_R -
    {\bf m^2_{\tilde u}}\, {\tilde u}_R^\dag\, \tilde u_R -
    {\bf m^2_{\tilde L}}\, {\tilde \ell}_L^\dag\, \tilde \ell_L - 
    {\bf m^2_{\tilde e}}\, {\tilde e}_R^\dag\, \tilde e_R -
    m^2_{\tilde L_5}\, {\tilde \ell}_{5L}^\dag\, {\tilde \ell}_{5L}\\ &\ -
    m^2_{\tilde{\bar L}_5}\, {\tilde \ell}_{5R}^\dag\, {\tilde \ell}_{5R} -
    m^2_{\tilde D_5}\, {\tilde d}_{5R}^\dag\, {\tilde d}_{5R} -
    m^2_{\tilde{\bar D}_5}\, {\tilde d}_{5L}^\dag\, {\tilde d}_{5L} -
    m^2_{\tilde N}\, {\tilde N}_L^\dag\, {\tilde N}_L -
    m^2_{\tilde{\bar N}}\, {\tilde N}_R^\dag\, {\tilde N}_R -
    \Big[{\bf m^2_{\tilde L_5,\tilde L}}\, {\tilde \ell}_{L}^\dag\, {\tilde \ell}_{5L}
     + {\rm h.c.}\Big]\\ &\  -
    \Big[{\bf m^2_{\tilde D_5, \tilde D}}\, {\tilde d}_{R}^\dag\, {\tilde d}_{5R}
     + {\rm h.c.}\Big] \ ,
\esp\ee
where $m^2_i$ represent the various mass parameters, in flavour space.
Moreover, the superpotential-induced soft
terms are written as
\be \label{eq:softW}\bsp
  {\cal L}_{\rm soft}^{(W)} =&\ \bigg[ - b\, h_u\cdot h_d\, +
    {\mathbf T_d}\, {\tilde d}^\dag\, {\tilde q}\cdot h_d\, +
    {\mathbf T_e}\, {\tilde e}^\dag\, {\tilde l}\cdot h_d\, -
    {\mathbf T_u}\, {\tilde u}^\dag\, {\tilde d}\cdot h_u\,
   - b_D\, \tilde d_{5R}^\dag\, \tilde d_{5L} \,
   - b_L\, \tilde\ell_{5R}^\dag \cdot \tilde\ell_{5L}\,
   - b_N\, \tilde N_R^\dag\, \tilde N_L\,\\&\
   - a_{k_N}\, \tilde N_R^\dag\, \tilde\ell_{5L}\cdot h_u\,
   + a_{h_N}\, \tilde N_L\, \tilde\ell_{5R}^\dag\cdot h_d\,
   + {\bf a_{\varepsilon_D}}\, \tilde d_{5R}^\dag\, \tilde q_L\cdot h_d\,
   + {\bf a_{\varepsilon_E}}\, \tilde e_R^\dag\, \tilde \ell_{5L}\cdot h_d\,
   - {\bf a_{\varepsilon_N}}\, \tilde N_R^\dag\, \tilde\ell_L\cdot h_u\,\\&\
   - {\bf b_{\kappa_D}}\, \tilde d_R^\dag \tilde d_{5L}\,
   - {\bf b_{\kappa_L}}\, \tilde\ell_{5R}^\dag\cdot\tilde\ell_L
   + {\rm h.c.} \bigg] \ ,
\esp\ee
where the first four terms are the usual MSSM soft terms, $b$ denotes the
bilinear Higgs interaction strength and ${\bf T_i}$ the various squark-Higgs
trilinear interactions in flavour space. The $a_{k_N}$ and $a_{h_N}$ parameters
represent the trilinear couplings of the vector-like sneutrinos and sleptons to
the Higgs fields, whilst the ${\bf a_{\varepsilon_i}}$ and ${\bf b_{\kappa_i}}$
parameters are three-dimensional vectors describing the mixing of the
vector-like and MSSM scalars. Similarly to their superpotential term counterparts, the
latter will be assumed vanishing for the first two generations, and
small for the third generation.

Once electroweak symmetry is broken, all particles with the same electric charge and lying in the same colour and spin
representations mix. The neutralino sector is identical to the MSSM one, consisting of four Majorana fermions $\chi_i^0$, $i = 1, \ldots, 4$,
which are linear combinations of the four neutral gaugino and higgsino gauge
eigenstates.
The lightest neutralino, $\chi_1^0$ is the first dark matter candidate we shall consider. 
The model also contains five physical neutrinos which are admixtures of the usual MSSM neutrinos with the fermionic neutral components
of $L_5$ and $\bar{L}_5$ as well as with the fermionic components of the gauge singlet 
chiral superfields $N$ and $\bar{N}$. The two heaviest exotic states are not
stable, since they can decay through the Yukawa-like  ${\bf \varepsilon_E}$ and
${\bf \varepsilon_N}$ terms in Eq.\eqref{eq:superpotential}, and thus they cannot be
 potential dark matter candidates. The second dark matter candidate considered 
is  in the sneutrino sector, which in the LND model consists of seven
physical scalars. The sneutrino mixing is described by the symmetric mass matrix
$M_{\tilde \nu}^2$. In the $(\tilde\nu_{5L}, \tilde N_L, \tilde \nu_e, \tilde
\nu_\mu, \tilde\nu_\tau, \tilde \nu_{5R}^\dag, \tilde N_R^\dag)$ basis, 
 matrix elements in the $7 \times 7$ symmetric matrix include soft terms, supersymmetric contributions as
well as $D$-term contributions and are given by
\be\bsp
  \big(M_{\tilde \nu}^2\big)_{11} = &\
    \frac18 \bigg[\frac{e^2}{s_w^2}+\frac{e^2}{c_w^2}\bigg]
    \bigg[v_d^2-v_u^2\bigg] + \mu_L^2 + m_{\tilde L_5}^2 + \frac12 k_N^2 v_u^2
    \ , \\
  \big(M_{\tilde \nu}^2\big)_{12} =&\
    \frac{1}{\sqrt{2}} \bigg[h_N \mu_L v_d + k_N \mu_N v_u\bigg]\ , \\
  \big(M_{\tilde \nu}^2\big)_{1(2+f)} = &\
    \mu_L \big({\bf \kappa_{L}}\big)_f +
    \frac12 k_N v_u^2 \big({\bf \varepsilon_{N}}\big)_f +
    \big({\bf m^2_{\tilde L_5,L}}\big)_f \ , \\
  \big(M_{\tilde \nu}^2\big)_{16} = &\ b_L\ , \\
  \big(M_{\tilde \nu}^2\big)_{17} = &\ \frac{1}{\sqrt{2}}
    \bigg[ a_{k_N} v_u - \mu k_N v_d \bigg] \ ,\\
  \big(M_{\tilde \nu}^2\big)_{22} = &\
    \mu_N^2 + m_{\tilde N}^2 + \frac12 h_N^2 v_d^2 \ , \\
  \big(M_{\tilde \nu}^2\big)_{2(2+f)} = &\
     \mu_N v_u \big({\bf \varepsilon_{N}}\big)_f +
     h_N v_d \big({\bf \kappa_{L}}\big)_f\bigg]\ ,\\
  \big(M_{\tilde \nu}^2\big)_{26} = &\ \frac{1}{\sqrt{2}}
    \bigg[ a_{h_N} v_d - \mu h_N v_u \bigg] \ ,\\
  \big(M_{\tilde \nu}^2\big)_{27} = &\ b_N\ , \\
  \big(M_{\tilde \nu}^2\big)_{(2+f)(2+f')} = &\
    \frac18 \bigg[\frac{e^2}{s_w^2}+\frac{e^2}{c_w^2}\bigg]
    \bigg[v_d^2-v_u^2\bigg]\delta_{ff'} + \big({\bf m_{\tilde L}^2}\big)_{ff'} +
     \frac12 v_u^2 \big({\bf \varepsilon_{N}}\big)_f
     \big({\bf \varepsilon_{N}}\big)_{f'} + \big({\bf \kappa_{L}}\big)_f
    \big({\bf \kappa_{L}}\big)_{f'} \ , \\
  \big(M_{\tilde \nu}^2\big)_{(2+f)7} = &\ \frac{1}{\sqrt{2}} \bigg[
    v_u \big({\bf a_{\varepsilon_N}}\big)_f -
    \mu v_d \big({\bf \varepsilon_{N}}\big)_f \bigg]\ , \\
  \big(M_{\tilde \nu}^2\big)_{66} = &\ \frac18
    \bigg[\frac{e^2}{s_w^2}+\frac{e^2}{c_w^2}\bigg]
    \bigg[v_u^2-v_d^2\bigg] + \mu_L^2 + m^2_{\tilde{\bar L}_5} +
    \frac12 h_N^2 v_d^2 + \sum_{f=1}^3 \big({\bf \kappa_{L}}\big)_f^2 \ , \\
  \big(M_{\tilde \nu}^2\big)_{67} = &\ \frac{1}{\sqrt{2}} \bigg[
    h_N v_d \mu_N + k_N v_u \mu_L + v_u
    \sum_{f=1}^3 \big({\bf \kappa_{L}}\big)_f \big({\bf \varepsilon_{N}}\big)_f
    \bigg]\ , \\
  \big(M_{\tilde \nu}^2\big)_{77} = &\ \mu_N^2 + m^2_{\tilde{\bar N}} +
    \frac12 k_N^2 v_u^2 + \frac12 v_u \sum_{f=1}^3
    \big({\bf \varepsilon_{N}}\big)_f^2 \ ,
\esp\ee
where $f, f'= 1$, 2 and 3 are generation indices, $\big(M_{\tilde \nu}^2\big)_{ij}=\big(M_{\tilde \nu}^2\big)_{ji}$  and all other elements  vanish.

\section{Parameter Space Exploration}
\label{sec:scan}
\subsection{Parameter space}
\label{sec:pspace}

As indicated in the Lagrangian introduced in the previous section, the LND model
parameter space is defined from a  large set of beyond the SM free
parameters. Assuming unification conditions and relying on the minimisation of
the scalar potential, this list can be further reduced. In the following, we
define the range of value allowed for each parameter relevant for our study.
We have verified that wider ranges did not yield any new
phenomenology. We fixed the values of all input parameters at the
supersymmetry-breaking scale, with the exception of the common MSSM sfermion and
Higgs mass $M_0$ and their common soft trilinear coupling $A_0$ that are defined
at the GUT-scale. This particular choice allows
us to analyse a large set of different scalar spectra, concentrating in particular on scenarios with 
light electroweak scalars and heavier strongly-interacting ones (that are
only marginally relevant for our dark matter study).  For the
supersymmetry-breaking scale we have taken the geometric mean of the masses of the lightest
($M_{\tilde u_1}$) and heaviest ($M_{\tilde u_6}$) up-type squarks,
$M_{\rm SUSY} \sim \sqrt{M_{\tilde u_1} M_{\tilde u_6}}$, and restricted it to
be smaller than 5~TeV.

We start with the superpotential parameters appearing in
Eq.~\eqref{eq:superpotential}. While the SM first and second generation Yukawa
couplings are neglected, we fix the third generation ones to the value
given in the Particle Data Group Review~\cite{Olive:2016xmw}. All other
parameters are left free and will be scanned over, with the exception of the
off-diagonal Higgs mixing parameter $\mu$ whose absolute value is fixed from the
scalar potential minimisation conditions. The supersymmetric masses of the
three pairs of vector-like supermultiplets $\mu_D$, $\mu_L$ and $\mu_N$ are
taken as varying in the GeV - TeV range,
\be
  \mu_D \in [1, 8]~{\rm TeV} \ ,\qquad
  \mu_L \in [0, 3]~{\rm TeV} \qquad\text{and}\qquad
  \mu_N \in [0, 5]~{\rm TeV} \ ,
\ee
whilst the vector-like Yukawa couplings are taken of ${\cal O}(1)$,
\be
  k_N \in [-1, 1] \qquad\text{and}\qquad h_N \in [-1, 1] \ .
\ee
As previously stated, we forbid any mixing between the vector-like sector and the
first two SM generations, so that the supersymmetric mass mixing parameters and
${\bf \varepsilon}$ Yukawa couplings solely involve the third
generation,
\be
  {\bf \kappa_D} = \bpm 0 \\ 0 \\ \kappa_D\epm  \ ,\qquad
  {\bf \kappa_L} = \bpm 0 \\ 0 \\ \kappa_L\epm  \ ,\qquad
  {\bf \varepsilon_D} = \bpm 0 \\ 0 \\ \varepsilon_D \epm  \ ,\qquad
  {\bf \varepsilon_E} = \bpm 0 \\ 0 \\ \varepsilon_E \epm \qquad\text{and}\qquad
  {\bf \varepsilon_N} = \bpm 0 \\ 0 \\ \varepsilon_N \epm  \ ,
\ee
with
\be
  \kappa_D \in [-10^{-6}, 10^{-6} ]~{\rm TeV} \ , \quad
  \kappa_L \in [-1, 1]~{\rm TeV}\ , \quad
  \varepsilon_N \in [-0.1, 0.1]\ , \quad
  \varepsilon_E \in [-1,1] \quad\text{and}\quad
  \varepsilon_D \in [-5,5]\times10^{-3}  \ .
\ee
The parameters in the down-type quark sector are restricted to be small by flavour constraints. The quoted intervals have been determined after scanning over larger ranges and restricting the parameters responsible for flavour-changing effects in the down-type quark and charged lepton sectors according to the constraints in Table~\ref{tab:constraints}.
The soft gaugino mass terms of Eq.~\eqref{eq:softino} are allowed to vary
independently in the GeV - multi-TeV range,
\be
  M_1 \in [0, 2]~{\rm TeV} \ , \qquad
  M_2 \in [0, 3]~{\rm TeV} \qquad\text{and}\qquad
  M_3 \in [0, 4]~{\rm TeV} \ ,
\ee
whilst the MSSM squark and Higgs mass parameters appearing in
Eq.~\eqref{eq:softscal} are imposed to unify to a common $M_0$ value,
\be
  M_0 \in [ 0, 5]~{\rm TeV} \ .
\ee
The extra squark mass and mixing parameters are chosen to vary 
independently,
\be
  m_{{\tilde D}_5}      \in [0 ,5]~{\rm TeV}\ , \qquad
  m_{\tilde{\bar D}_5} \in [0 ,5]~{\rm TeV}\ , \qquad
 {\bf m^2_{\tilde D_5, \tilde D}} =
    \bpm 0 \\ 0 \\ m^2_{\tilde D_5, \tilde D}\epm \quad\text{with}\quad
    m^2_{\tilde D_5, \tilde D} \in [-0.1, 0.1]~{\rm TeV}^2 \ .
\ee
These new states introduce some of the specific features of the LND model, but our dark matter analysis and collider signals are largely independent of this choice.
By contrast, the mass
parameters of the extra sneutrinos and sleptons are also specific parameters
of the model, but very relevant for what concerns cosmology.  We 
keep all of them free and allow them to vary independently, again in the
multi-TeV range,
\be
  m_{\tilde N}          \in [0 ,6]~{\rm TeV}\ , \qquad
  m_{\tilde{\bar N}}     \in [0 ,6]~{\rm TeV}\ , \qquad
  m_{{\tilde L}_5}      \in [0 ,1]~{\rm TeV}\qquad\text{and}\qquad
  m_{\tilde{\bar L}_5} \in [0 ,1]~{\rm TeV}\ ,
\ee
and we fix the mass mixing between the MSSM and the vector-like sleptons as
\be
  {\bf m^2_{\tilde L_5,\tilde L}} = \bpm 0\\0\\ m^2_{\tilde L_5,\tilde L}\epm
  \quad\text{with}\quad m^2_{\tilde L_5,\tilde L} \in [-0.1, 0.1]~{\rm TeV}^2\ .
\ee
The rest of the soft parameters involving the MSSM sfermions and Higgs bosons
are assumed to unify at the GUT scale, so that
all squark trilinear couplings to the Higgs sector are set to a common $A_0$
value multiplied by the relevant SM fermion Yukawa coupling. In addition, all trilinear
scalar couplings involving two vector-like sfermions are taken vanishing.
Moreover, all
bilinear terms are fixed to a common $B_0$ value, with the exception of the Higgs
mixing parameter $b$ whose value is driven by the scalar potential minimisation.
We thus choose
\be
  B_0\in [-5, 5]~{\rm TeV} \ , \qquad
  A_0\in [-2, 2] \qquad\text{and}\qquad
  a_{k_N} = a_{h_N} = 0 \ .
\ee
As for all other interactions involving the mixing of a vector-like and an MSSM
particle, we enforce the soft ones to vanish for  the first two generations,
\be
  {\bf a_{\varepsilon_D}} = \bpm 0\\0\\ a_{\varepsilon_D}\epm \ , \qquad
  {\bf a_{\varepsilon_E}} = \bpm 0\\0\\ a_{\varepsilon_E}\epm
  \qquad\text{and}\qquad
  {\bf a_{\varepsilon_N}} = \bpm 0\\0\\ a_{\varepsilon_N}\epm \ ,
\ee
and, for simplicity, fix the input values of the three remaining free parameters
to zero,
\be
  a_{\varepsilon_D} = a_{\varepsilon_E} = a_{\varepsilon_N}  = 0 \ .
\ee
Since the SM Higgs-boson mass is taken equal to its measured value, the Higgs
sector is fully defined from the conditions stemming from the minimisation of
the scalar potential, once one extra parameter is fixed,  as all parameters
contributing at the one-loop order and beyond are already defined above. The
ratio of the vacuum expectation values of the neutral components of the two
Higgs doublets, $\tan\beta$,  is allowed to vary in the following range,
\be
  \tan\beta \in [1, 60] \ .
\ee
The parameter space is now defined by the 25 new physics parameters and
one sign is listed in Table~\ref{tab:scan_lim}, where we summarise  the free parameters and
 the range over they are scanned.

\begin{table}
  \setlength\tabcolsep{7pt}
  \renewcommand{\arraystretch}{1.4}
  \begin{tabular}{c|c||c|c}
    Parameter      & Scanned range& Parameter      & Scanned range\\
    \hline
    $\mu_D$          & $[1,8]$~TeV
         & $M_1$     & $[0,2]$~TeV\\
    $\mu_N$          & $[0,5]$~TeV
         & $M_2$     & $[0,3]$~TeV\\
    $\mu_L$          & $[0,3]$~TeV
         & $M_3$     & $[0,4]$~TeV\\
    ${\rm sgn}(\mu)$ &$\pm 1$
         & $m_{{\tilde L}_5}$, $m_{\tilde{\bar L}_5}$ & $[0,1]$~TeV\\
    $h_N$, $k_N$     & $[-1,1]$
         & $m_{{\tilde D}_5}$, $m_{\tilde{\bar D}_5}$ & $[0,5]$~TeV\\
    $\kappa_L$       &$[-1,1]$~TeV
         & $m_{{\tilde N}}$, $m_{\tilde{\bar N}}$ & $[0,6]$~TeV\\
    $\kappa_D$       &$[-1,1]\times10^{-6}$~TeV
         & $m^2_{\tilde L_5,\tilde L}$, $m^2_{\tilde D_5, \tilde D}$
         &$ [-0.1,0.1]$~TeV$^2$\\
    $\varepsilon_N$  &$[-0.1,0.1]$
         &$A_0$ &$[-2,2]$~TeV\\
    $\varepsilon_E$  &$[-1,1]$
         & $B_0$ &$[-5,5]$~TeV   \\
    $\varepsilon_D$  &$[-5,5] \times 10^{-3}$
         & $\tan\beta$    & $[1, 60]$ \\
    $M_0$  &$ [0,5]$~TeV
      \end{tabular}
   \caption{\label{tab:scan_lim} \it Range of the free parameters of the
    model scans.  The SM parameters are fixed to the values reported in
    the Particle Data Group Review~\cite{Olive:2016xmw} and 
    all non-listed parameters are fixed to zero.}
\end{table}

\subsection{Analysis setup and experimental constraints}
\label{sec:tools}

\begin{table}
  \setlength\tabcolsep{7pt}
  \renewcommand{\arraystretch}{1.6}
  \begin{tabular}{l|c||l|c}
    Observable & Constraint& Observable & Constraint\\
    \hline
    BR$(B^0\to X_s\gamma)$ & $[2.99, 3.87]\times10^{-4}$ \cite{Amhis:2016xyh} &
       ${\rm BR}(B\to\tau\nu_\tau) / {\rm BR}_{SM}(B\to\tau\nu_\tau)$ &
       $[0.15,2.41]$ \cite{Asner:2010qj}\\
    BR$(B^0_s\to\mu^+\mu^-)$ & $[1.1, 6.4]\times10^{-9}$ \cite{Aaij:2012nna}&
      BR$(\tau\to e\gamma)$  & $[0, 3.3] \times 10^{-8}$ \cite{Aubert:2009ag}\\
    BR$(\tau \to \mu \gamma)$ &$  [0,4.4] \times 10^{-8}$ \cite{Aubert:2009ag}&
      BR$(\tau \to e \pi) $ &$  [0,8.0] \times 10^{-8}$ \cite{Miyazaki:2007jp}\\
    BR$(\tau\to\mu\pi)$ & $[0, 1.1] \times 10^{-7}$ \cite{Miyazaki:2007jp} &
      BR$(\tau\to3\mu)$ & $[0, 2.1] \times 10^{-8}$ \cite{Aad:2016wce}\\
    $\Delta M_s$ & $[10.2,26.4]$~ps$^{-1}$ \cite{Jubb:2016mvq} &
      $\Delta M_d $ & $[0.29,0.76]$~ps$^{-1}$ \cite{Jubb:2016mvq}\\
    BR$(Z\to e\mu)$ & $[0,7.5]\times 10^{-7}]$ \cite{Aad:2014bca} &
      BR$(h \to e \mu)$ &$[0,3.5]\times 10^{-4}$ \cite{Khachatryan:2016rke}\\
    EWPO tests& $ \leq 2\sigma$ \cite{Baak:2014ora,Eriksson:2009ws,%
      Ilnicka:2015jba} & $\chi^2(\hat\mu)$ & $\leq 111.6$ \\
  \end{tabular}
  \vspace*{.2cm}
  \begin{tabular}{l|c||l|c}
    Mass & Constraint & Mass & Constraint \\
    \hline
    $M_{\chi^0_2}$     & $>62.4$ GeV  \cite{Olive:2016xmw} &
    $M_{\chi^\pm_1}$   & $>103.5$ GeV \cite{Olive:2016xmw}\\
    $M_{\chi^0_3}$     & $>99.9$ GeV  \cite{Olive:2016xmw} &
    $M_{\tilde{e}}$    & $>107$ GeV   \cite{Olive:2016xmw}\\
    $M_{\chi^0_4}$     & $>116 $ GeV  \cite{Olive:2016xmw} &
    $M_{\tilde{\mu}}$  & $>94$ GeV    \cite{Olive:2016xmw}\\
    $M_{\tilde{g}}$    & $>1.75$ TeV  \cite{Khachatryan:2016xvy} &
    $M_{\tilde{\tau}}$ & $>81$ GeV    \cite{Olive:2016xmw}\\
    $M_{\tilde{t}}$    & $>750$ GeV   \cite{Aaboud:2016lwz} &
    $M_{\tau^\prime}$  & $>103$ GeV   \cite{Achard:2001qw}\\
    $M_h$ & $125.09\pm3$ GeV \cite{Chatrchyan:2012xdj} &&\\
  \end{tabular}
  \caption{\label{tab:constraints} \it Set of low-energy and flavour
    physics constraints  imposed within our LND model scanning
    procedure (upper) and mass bounds imposed on the Higgs boson and new physics
    states (lower).}
\end{table}

In order to explore the parameter space defined in Section~\ref{sec:pspace},
we have implemented the LND model in the \textsc{Sarah} 4.12.2 package~\cite{%
Staub:2013tta}, which we have used to generate the corresponding \textsc{SPheno}
(version 4.0.3) output~\cite{Porod:2011nf}. With this last code, we
derive the value of the model parameters at the electroweak scale through
their renormalisation group running from the input scale, and extract the
particle
spectrum. In order to assess the phenomenological viability of the different
scenarios probed during the scan, we enforce the compatibility with several
flavour, collider and low-energy physics observables calculated by {\sc SPheno}
and
summarised in Table~\ref{tab:constraints}. The scenarios considered in our study
must satisfy rare $B$-decay constraints~\cite{Amhis:2016xyh,Aaij:2012nna,%
Asner:2010qj},
\be
    {\rm BR}(B^0 \to X_s \gamma)   \in [2.99,3.87]\times10^{-4} \ , \quad
    {\rm BR}(B^0_s \to \mu^+\mu^-) \in [1.1, 6.4] \times10^{-9}
    \quad\text{and}\quad
    \frac{{\rm BR}(B \to \tau\nu_\tau)}{{\rm BR}_{\rm SM}(B \to \tau\nu_\tau)}
     \in [0.15,2.41]\ ,
 \ee
rare tau-decay constraints~\cite{Aubert:2009ag,Miyazaki:2007jp,Aad:2016wce},
\be\bsp
  &
    {\rm BR}(\tau \to e \gamma)    \in [0,   3.3] \times10^{-8}\ , \qquad
    {\rm BR}(\tau \to \mu \gamma)  \in [0,   4.4] \times10^{-8}\ , \qquad
    {\rm BR}(\tau \to e \pi)       \in [0,   8.0] \times10^{-8}\ , \\
  &\hspace*{3cm}
    {\rm BR}(\tau \to \mu \pi)     \in [0,   1.1] \times10^{-7}\ , \qquad
    {\rm BR}(\tau \to 3\mu)        \in [0,   2.1] \times10^{-8}\ ,
\esp\ee
$B$-meson oscillation constraints~\cite{Jubb:2016mvq},
\be
  \Delta M_s \in [10.2,26.4]~{\rm ps}^{-1} \ , \qquad
  \Delta M_d \in [0.29,0.76]~{\rm ps}^{-1}\ ,
\ee
and flavour-violating $Z$-boson~\cite{Aad:2014bca} and Higgs-boson~\cite{%
Khachatryan:2016rke} decay bounds,
\be
  {\rm BR}(Z\to e \mu) \in [0, 7.5]\times 10^{-7}
  \qquad\text{and}\qquad
  {\rm BR}(h \to e \mu)  \in [0, 3.5] \times 10^{-4} \ .
\ee
Moreover, we impose the compatibility with electroweak precision observables
(EWPO) at the $2\sigma$ level~\cite{Baak:2014ora}, using a correlation
function based on the oblique parameters~\cite{Eriksson:2009ws,Ilnicka:2015jba}.
Thanks to the interface of {\sc SPheno}
with {\sc HiggsBounds} version 4.3.1~\cite{Bechtle:2008jh} and {\sc
HiggsSignals} version 1.4.0~\cite{Bechtle:2013xfa}, we verify the
consistency of the Higgs sector with experimental measurements of LHC Run~1. In
practice, we check that the Higgs-boson mass, gluon and vector-boson fusion
production cross sections (computed with the \textsc{SusHi} program version
1.5~\cite{Harlander:2012pb}) and signal strengths agree with data up to  
 deviations corresponding to a global $\chi^2(\hat\mu)$ quantity of at most 111.6, which corresponds to a $2\sigma$ level of agreement for the number of considered observables.

Additionally, we constrain superpartners masses (of the MSSM sector) from direct search bounds~\cite{Olive:2016xmw,%
Khachatryan:2016xvy,Aaboud:2016lwz}. We impose
that the gluino mass ($M_{\tilde g}$), the neutralino and chargino masses
($M_{\tilde\chi^0_i}$ and $M_{\tilde\chi^\pm_i}$), the slepton masses
($M_{\tilde e}$, $M_{\tilde\mu}$ and $M_{\tilde\tau}$) and the stop mass
($M_{\tilde t}$) satisfy
\be\bsp
  M_{\tilde g} > 1.75~{\rm TeV} \ , \quad
  M_{\tilde{\chi}^0_2} > 62.4~{\rm GeV} \ , \quad
  M_{\tilde{\chi}^0_3} > 99.9~{\rm GeV} \ , \quad
  M_{\tilde{\chi}^0_4} > 116~{\rm GeV} \ , \quad
  M_{\tilde{\chi}^\pm_1} > 103.5~{\rm GeV} \ , \\
  M_{\tilde e} > 107~{\rm GeV} \ , \quad
  M_{\tilde \mu} >  94~{\rm GeV} \ , \quad
  M_{\tilde \tau} >  81~{\rm GeV} \ , \quad
  M_{\tilde t} > 750~{\rm GeV} \ ,
\esp\ee
and enforce the vector-like lepton mass $M_{\tau'}$ to obey the LEP
bound~\cite{Achard:2001qw},
\be
  M_{\tau '} > 103~{\rm GeV}.
\ee
We now proceed to our analysis.  
We perform a scan of the parameter space by relying on the
Metropolis-Hastings sampling method \cite{1970Bimka..57...97H} in which
the model free parameters vary as in Table \ref{tab:scan_lim} and are restricted by
 the constraints of Table \ref{tab:constraints},
with the additional requirement that the LSP has to be neutral. For each point,
we perform the dark matter analysis with \textsc{micrOMEGAs}
version 4.3.1~\cite{Belanger:2014vza}, which allows us to calculate all DM
observables used in the analysis of Section~\ref{sec:DM} from the LND
\textsc{CalcHep}~\cite{Belyaev:2012qa} model file
generated by {\sc Sarah}. In Section~\ref{sec:detector}, we perform a collider
analysis of a few benchmark scenarios representative of the different spectra
favoured by cosmology, by relying on the {\sc MG5\_aMC@NLO}~\cite{Alwall:2014hca}
platform and an LND UFO model file~\cite{Degrande:2011ua} generated by {\sc
Sarah}. The interfacing of the various programmes and
our numerical analysis have been performed with a modified version of the
\textsc{pySLHA} package version 3.1.1~\cite{Buckley:2013jua}.


\section{Dark matter phenomenology}\label{sec:DM}

Having presented our model, the leading experimental constraints that we
subject it to, and the methodology that we use in order to explore and assess the
viability of the parameter space, we now proceed to present the results of our dark matter analysis. We
divide the discussion into two parts, depending on the nature of the dark matter
candidate (neutralino or sneutrino).

\subsection{Neutralino dark matter}
\label{sec:neutralino}

The lightest neutralino has, since long, been the most celebrated dark matter candidate of the MSSM. However, in the MSSM, barring co-annihilations and funnels, the possibilities for neutralino dark matter with an ${\cal{O}}(10^2)$ GeV mass are now severely constrained.
In particular, almost pure higgsinos and winos tend to be under-abundant, unless
their mass lies above about 1~TeV, as the cosmologically-attractive possibility
of a pure higgsino with a mass below the $W$-boson mass $M_W$ is excluded from
chargino searches at LEP~\cite{Heister:2002mn,Abdallah:2003xe,Acciarri:1999km,%
Abbiendi:2003sc}. Binos, on the other hand, tend to be overabundant by a few factors, 
unless either they can annihilate through the $t$-channel exchange of a
sufficiently light sfermion into SM fermions, or they contain a substantial
higgsino or wino fraction. The former case is disfavoured by sfermion searches
at the LHC, whereas direct detection experiments~\cite{Aprile:2017iyp} strongly
constrain the mixed bino-higgsino scenario. The mixed bino-wino case is less constrained and constitutes one of the remaining possibilities
for sub-TeV natural neutralino dark matter\footnote{Natural in the sense of how rapidly
the predicted dark matter abundance changes with small variations of the model
parameters~\cite{Perelstein:2011tg,Perelstein:2012qg,Belanger:2014vua}.}. We
refer to Ref.~\cite{Profumo:2017ntc} for a recent detailed account of existing
constraints.

\begin{figure}
 \centering
 \includegraphics[scale=0.6]{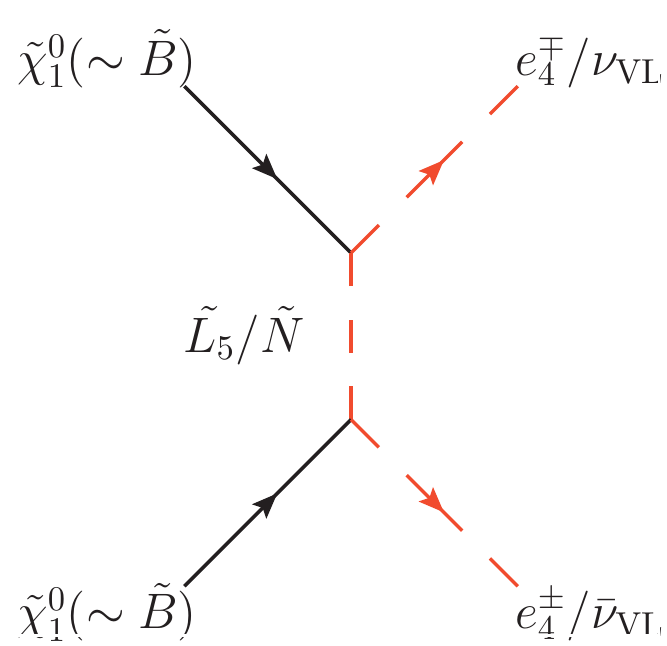}
 \caption{\it Representative dark matter annihilation diagram into vector-like fermions.
 \label{fig:neutann}}
\end{figure}

In the context of MSSM extensions with vector-like fermions, however, the possibility
of an almost pure bino dark matter with a mass of up to a few hundreds of GeV
can be viable~\cite{Abdullah:2015zta,Abdullah:2016avr}. In such
scenarios, binos can annihilate into vector-like fermions through the
$t$-channel exchange of the corresponding sfermion as illustrated in
Figure~\ref{fig:neutann}. In the QUE and QDEE models, these annihilation
channels can constitute an efficient-enough mechanism for depleting binos in the
early Universe due to the interplay of two effects. First, binos can annihilate
into vector-like weak-singlet leptons $E$ carrying hypercharge $Y_E = 1$. Since
the sfermion-mediated annihilation cross section of binos into fermions scales
as $Y^4$, this annihilation channel is particularly enhanced. Second, although $SU(2)_L$
singlet fermions also exist in the MSSM, the corresponding annihilation cross section is suppressed by the masses of the light SM fermions. The vector-like fermions present in MSSM extensions, on the other hand, are (necessarily) much heavier, hence this suppression is no longer present. It is the interplay of these two factors that renders heavier bino dark matter a viable option in the QUE and QDEE models, making it possible to achieve masses as high as $\sim 450$ GeV in the former case and $\sim 600$ GeV in the latter \cite{Abdullah:2015zta,Abdullah:2016avr}.
In the LND model, however, the situation is slightly different. Although, in
this model too the binos can annihilate into heavy electron (and neutrino) pairs
through the $t$-channel exchange of the corresponding sfermions, now the new
leptons belong to an
$SU(2)_L$ doublet with an hypercharge $Y_L = 1/2$. This implies that, all other parameters being identical, the bino annihilation cross section is suppressed by a factor $1/16$ relatively to the QUE and QDEE models. We thus expect the phenomenology of neutralino dark matter to be more similar to the MSSM one than that of the other two GUT-inspired MSSM extensions with vector-like fermions.

\begin{figure}
  \centering
  \includegraphics[width=0.49\columnwidth]{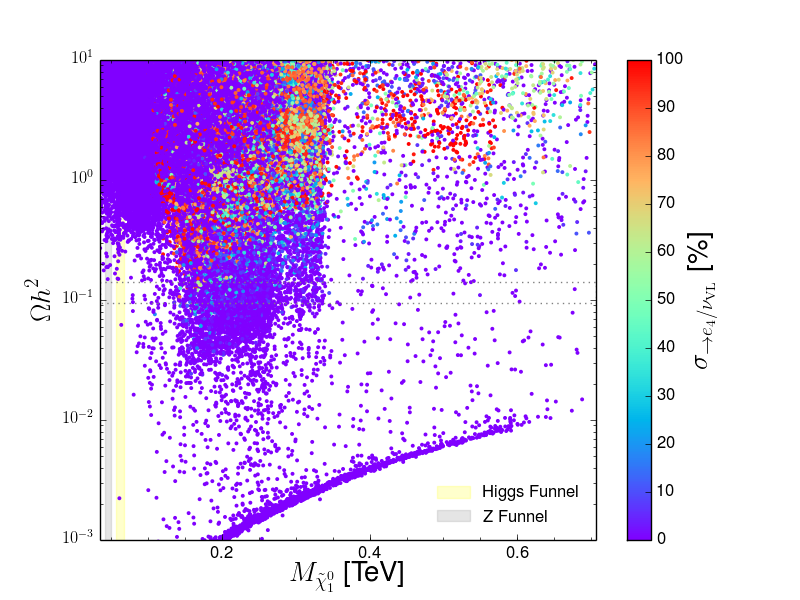}
  \includegraphics[width=0.49\columnwidth]{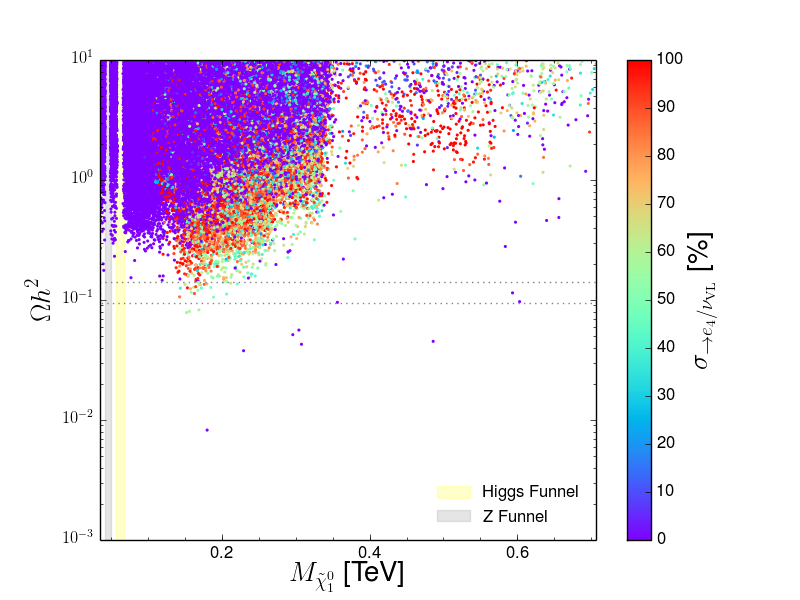}
  \caption{\it Relic abundance of neutralino dark matter in the LND model as a
    function of the dark matter mass, with all known funnels and
    co-annihilation channels (left)  or without those channels (right). The
    different colours correspond to the relative importance of the contributions
    from annihilations into vector-like fermions to the total dark matter
    annihilation cross section (blue to red for increasing contribution). The
    horizontal dashed lines indicate the region favoured by Planck
    data~\cite{Ade:2015xua}.
  \label{fig:neutchannels}}
 \end{figure}

In Figure \ref{fig:neutchannels} we present the neutralino relic abundance as a
function of the dark matter mass, highlighting in different colours the contribution of these novel annihilation channels to the total dark matter annihilation cross section. In the left panel we show all the
scenarios probed in our scanning procedure that respect the constraints
described in Section~\ref{sec:tools}, whereas in the right panel we exclude scenarios in which dark matter depletion is dominated by funnels ($Z/h/H/A$) or co-annihilations. Several comments are in order.

First, we recover the well-known result that
sub-TeV higgsinos and winos lie below the Planck region (with $\Omega h^2 \sim
0.12$~\cite{Ade:2015xua}), {\it i.e.} the
predicted relic density is smaller by one or two orders of magnitude than the
observed one. This is illustrated on the bottom of the left subfigure as
 a line-like
accumulation of scenarios. In contrast, almost pure bino dark matter that does
not annihilate into vector-like fermions can be either overabundant or
under-abundant, depending on whether or not co-annihilations and funnels are
efficient in depleting DM. The blue parameter space points for which the Planck
measurements are exactly met correspond to scenarios of bino dark matter either
annihilating through a quasi on-shell $Z/h/H$ or $A$ boson, or co-annihilating
with MSSM sparticles. Such configurations are also present in the MSSM. The
novel feature appearing in the LND model are the red points, which
correspond exactly to situations in which binos annihilate into vector-like
leptons. While co-annihilations with the corresponding sfermions are also
possible, they are not necessary to reproduce the Planck measurements.

Second, we observe the existence of a lower bound in the predicted dark matter abundance for binos annihilating exclusively into vector-like leptons. This limit is due to the fact that the interactions involved in annihilation diagrams such as the one depicted in Figure \ref{fig:neutann} result from gauge couplings, which implies that their magnitude is essentially fixed. Then, in the absence of additional annihilation processes, these interactions can be efficient only up to a certain point in depleting dark matter, which corresponds to the observed lower bound in $\Omega h^2$. This lower limit scales roughly as the squared bino mass,
which is a consequence of the fact that, for large enough dark matter masses, $\left\langle \sigma v \right\rangle$ is roughly proportional to the inverse square of the dark matter mass, a dependence which reflects upon the predicted relic density. The situation is fairly similar to the scaling of the wino and/or higgsino abundance as a function of the neutralino mass. 

Third, the rather sharp cutoff observed in the red points around a mass of 100 GeV is simply due to the fact that the vector-like leptons (and, in particular, the heavy electrons) cannot be lighter than about 100~GeV, because of the experimental constraints on their mass discussed in Section \ref{sec:tools}.

So, as  anticipated, the neutralino dark matter phenomenology we recover is
fairly similar to the MSSM one. Due to the hypercharge suppression of processes such as the one depicted in Figure \ref{fig:neutann}, annihilation into vector-like fermions is not as efficient in the LND model as in the QUE and QDEE ones. It is, hence, not possible to reach bino masses larger than $\sim 200$ GeV while imposing all existing experimental constraints and explaining the observed dark matter density in the Universe.
Although the dark matter annihilation channels might differ drastically, the
accessible masses are at the end comparable to those that would be obtained in the MSSM. However, as we will see in Section \ref{sec:detector}, the existence of the vector-like (s)fermions can give rise to interesting, novel phenomenological signatures at the LHC and provide additional handles for collider dark matter searches.

\subsection{Sneutrino dark matter}
\label{sec:sneutrino}

\begin{figure}
  \centering
  \includegraphics[width=0.49\columnwidth]{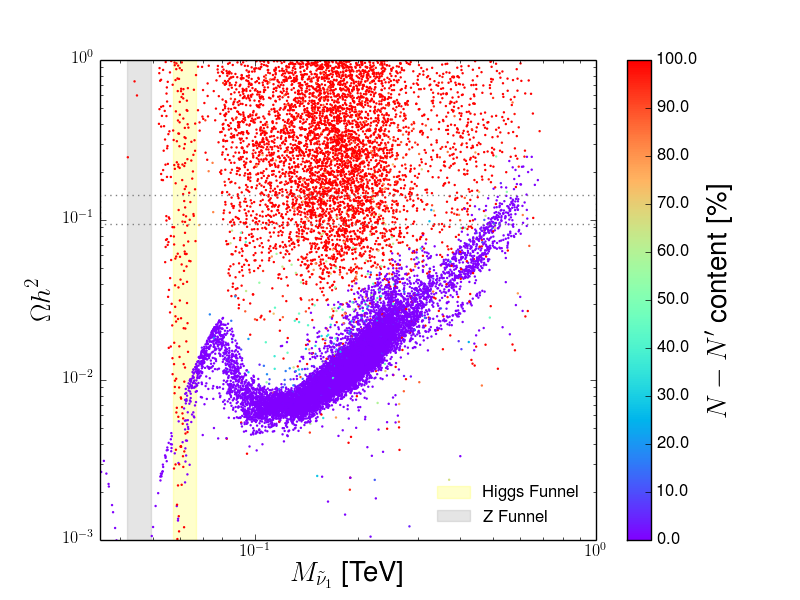}
  \includegraphics[width=0.49\columnwidth]{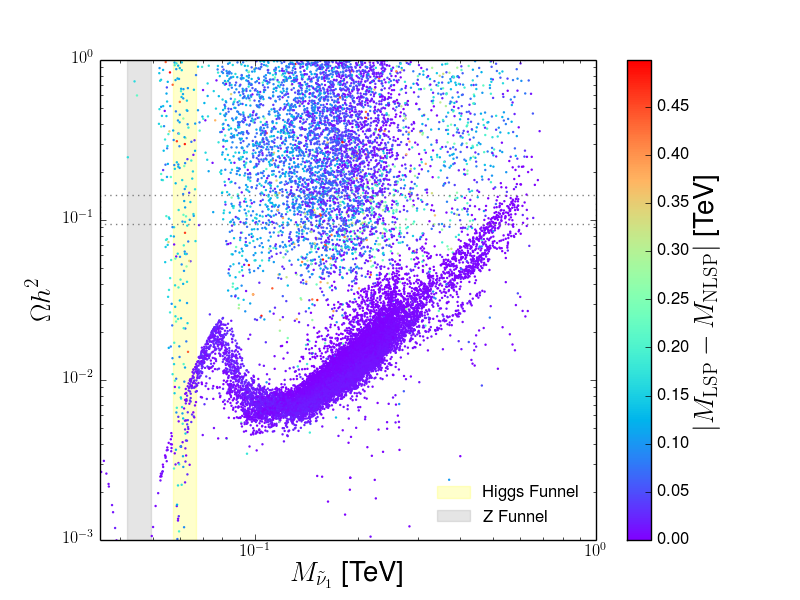}
  \caption{\it Sneutrino relic abundance as a function of its mass. In the left
    panel we highlight in blue the parameter space points for which the lightest
    sneutrino is dominated by a doublet component, whereas red points are
    essentially singlet-like. In the right panel, we indicate which
    of the parameter space points are characterised by large (red) or small
    (blue) mass splittings between the lightest sneutrino and the NLSP.
  \label{fig:sneutrelic}}
 \end{figure}

As already explained in Section \ref{sec:model}, LND sneutrinos can be a random admixture of the MSSM ($SU(2)$ doublet) left-handed sneutrinos and the vector-like left- and right-handed $SU(2)$ doublet or singlet ones. A first finding from our parameter space scan is that, as expected
(\textit{cf.~e.g.} Ref.~\cite{Arina:2007tm}), mostly doublet-like sneutrino dark matter can be perfectly compatible with the requirement to reproduce the observed dark matter abundance in the Universe, but is excluded by direct detection experiments due to the strong coupling to the $Z$ boson. This is a well-known feature in  the MSSM which persists in the LND model.
In order to illustrate it, in the left panel of Figure~\ref{fig:sneutrelic} we show the sneutrino relic abundance as a function of its mass, highlighting in different colours (red to blue) the increasing doublet fraction. For simplicity, we ignore scenarios with MSSM-like sneutrinos. We observe that mostly doublet-like scenarios (blue points) can satisfy the Planck constraint for sneutrino masses around $600 - 800$ GeV, a range which is comparable to the usual MSSM sneutrino dark matter scenario \cite{Arina:2007tm}. These scenarios are, nonetheless, found to be in severe conflict with direct detection constraints.

In principle, the presence of additional light leptonic doublets could also
provide the necessary contributions to tame down the discrepancy between the
measured and predicted values of the anomalous magnetic moment of the muon,
despite the fact that mixing is only allowed with the third generation of SM
fermions~\cite{Choudhury:2017fuu}. This appealing option turns out to be
strongly disfavoured by cosmology, so that one ends up with a situation similar
to the MSSM one. From now on, we will  not analyse further  doublet-like
sneutrino DM candidates.

Singlet-like scenarios (red points), on the other hand, offer much more freedom both from the viewpoint of the Planck-allowed sneutrino masses and as far as direct detection constraints are concerned. The abundance of singlet-like scenarios is determined through the interplay of several dark matter depletion processes including direct annihilations into Higgs boson pairs, annihilations through quasi-on-shell $s$-channel scalars and sfermion exchange, and co-annihilations.
The impact of the latter is in particular illustrated in the right panel of
Figure~\ref{fig:sneutrelic}, in which we highlight in different colours (red to blue) scenarios with  decreasing mass splitting between the lightest sneutrino and the NLSP and which indicate   increasing co-annihilation contributions. The observed relic abundance in the Universe can be reproduced for a large range of mass splittings, which implies that sneutrinos can be a cosmologically viable option with or without co-annihilations.

 \begin{figure}
 	\centering
	\includegraphics[width=0.49\columnwidth]{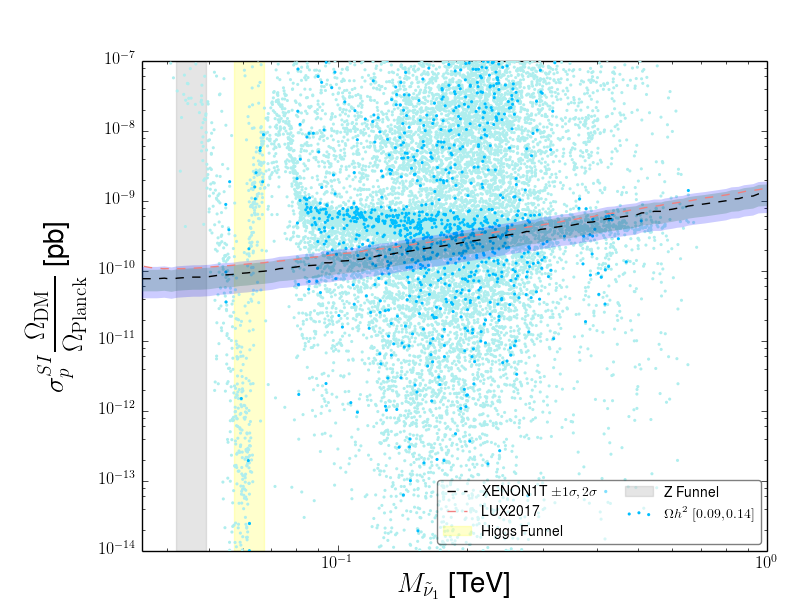}
	\includegraphics[width=0.49\columnwidth]{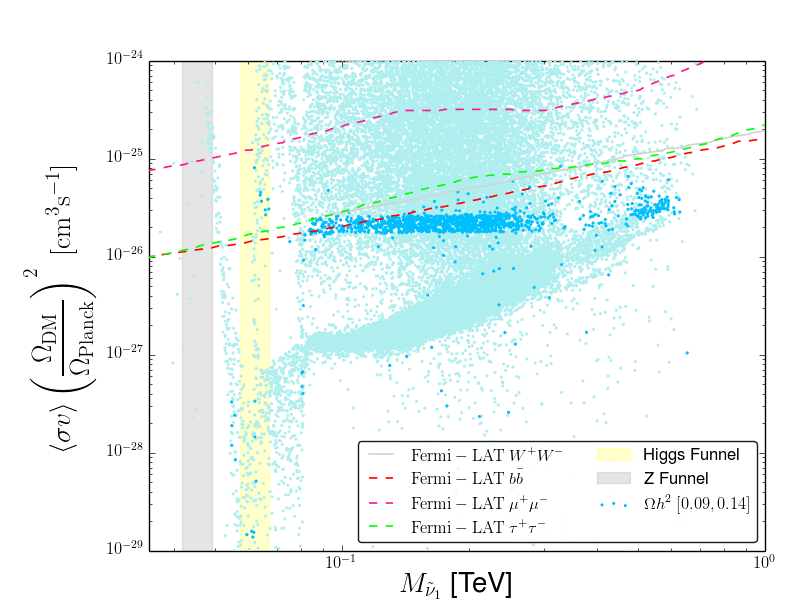}
	\caption{\it Direct (left panel) and indirect (right panel) detection constraints on sneutrino dark matter in the LND model as a function of the dark matter mass. In both cases we highlight, in darker blue, the parameter space points for which the two-sided Planck constraint can be satisfied.
	\label{fig:sneutddin}}
 \end{figure}

The impact of the DM direct and indirect detection constraints for sneutrino
dark matter on the LND model is shown in Figure~\ref{fig:sneutddin}.
In the left panel we present the DM-proton spin-independent scattering cross section against the sneutrino mass and compare it with the latest exclusion bounds from
LUX~\cite{Akerib:2015rjg,mark_mitchell_2016_61108} (red-dashed line) and
XENON1T~\cite{Aprile:2015uzo} (black-dashed line and shaded $3\sigma$ band). In order to account for the
possibility of sneutrinos comprising only a subleading dark matter component, the scattering cross section has been rescaled according to the predicted dark matter abundance for each scenario, which we assume to scale identically to its present-day local density, an we ignore configurations yielding over-abundant dark
matter.
In the right panel we instead show the predicted zero-velocity thermally
averaged self-annihilation cross section and compare it with the latest bounds
from the Fermi satellite mission~\cite{Ackermann:2015zua} for different
annihilation channels. Although the exact nature of the annihilation products
may vary substantially, the constraints for hadronic (quarks/gauge bosons)
channels tend to correspond more to the actual constraints on our scenarios. The
self-annihilation cross section has in this case been rescaled by
$\Omega_{\rm pred}^2/\Omega_{\rm Planck}^2$, again in order to account for the
possibility of under-abundant scenarios. In both panels, the parameter space points satisfying both the upper and the lower Planck bound are highlighted in darker blue and we omit points for which dark matter would be over-abundant.

As we can observe from the left panel of  Figure~\ref{fig:sneutddin}, the
parameter space points for which $M_{\tilde{\nu}_1} \sim M_h/2$ and which
satisfy the relic density constraint lie within a band that exhibits a rather
standard ``funnel''-like behaviour.
Below this mass value, efficient annihilation of singlet-like sneutrinos into light SM fermions requires rather large couplings to the Standard Model particles, especially the Higgs boson, which in turn implies that the corresponding scenarios are characterised by relatively large spin-independent scattering cross sections. This brings them in conflict with the recent LUX and XENON1T constraints, which are only satisfied if the sneutrinos annihilate through a quasi-resonant Higgs boson in the early Universe.
Once the $W$-boson mass threshold is crossed, we observe another abrupt drop in
$\sigma_p^{SI}$ since for $M_{\tilde{\nu}_1} > M_W$ the sneutrinos can annihilate directly into a pair of $W$ (and, eventually, $Z$) bosons. Still, direct detection constraints exclude most of the parameter space lying within this band for sneutrino masses roughly up to 200 GeV. For larger masses, the present-day sensitivity of direct detection experiments is no longer sufficient to exclude these scenarios.

Essentially the same structure is observed in the right panel of
Figure~\ref{fig:sneutddin}, without the sudden drop around $M_W$ since the total
dark matter self-annihilation cross section remains roughly constant in order to
satisfy the Planck bounds. Besides, the occasional scattered points that lie outside the main dark blue band correspond to scenarios with large co-annihilation contributions or to cases in which dark matter annihilates predominantly into Higgs boson pairs or vector-like leptons.
The larger spread of the dark blue points in the left panel of the figure with
respect to the right one is due to the fact that, with the exception of co-annihilation, indirect detection probes \textit{the same} processes that determine the dark matter abundance in the Universe. For instance, if dark matter annihilates predominantly into Higgs boson pairs, its direct detection prospects are rather modest whereas its indirect detection perspectives are almost identical to scenarios with a $WW$ final state, since the final annihilation products are similar and the total thermally averaged self-annihilation cross section is roughly the canonical one.

The DM detection perspectives of the model are hence good, since most of the parameter space that has not yet been excluded by direct or indirect detection lies within a factor of a few from current bounds.
We can expect that at least the most ``canonical'' scenarios will be probed
with in the next decade or so.
However, LND sneutrino dark matter candidates
behave rather similarly to usual sneutrino dark matter in the MSSM augmented with a right-handed neutrino chiral supermultiplet. Most of the features discussed here are present in this scenario too, with the most important differences coming from the existence of a few additional annihilation channels (the contribution of which we have, however, found to be rather modest) and the presence of a handful of additional co-annihilation channels. Both of these features do not alter the global picture of sneutrino dark matter with respect to more conventional scenarios.
Despite this, we should remind that given the current experimental constraints, essentially all neutralino dark matter scenarios necessitate a $\mu$ parameter that lies at the TeV scale or above. As it has been pointed out~\cite{Perelstein:2011tg,Perelstein:2012qg,Belanger:2014vua}, since the $\mu$ parameter is related to the $Z$-boson mass already at tree-level, this introduces high levels of fine-tuning pushing the theory towards unnatural territories. In sneutrino dark matter scenarios such as the one we just studied, the $\mu$ parameter is decoupled from the dark matter mass and can be fixed much closer to the $Z$-boson mass as required by naturalness. In this respect, the presence of a second dark matter candidate in the form of the (mostly singlet-like) sneutrino in the LND model constitutes of an interesting novelty both with respect to the MSSM and to the QUE and QDEE setups.

\section{Prospects at the HL-LHC}
\label{sec:detector}
The results presented in the previous sections show that there exist LND
configurations compatible with cosmological constraints as well as with flavour,
Higgs and low-energy physics observables. We therefore single out several
representative benchmark scenarios to  study the corresponding LHC
phenomenology in more depth, and turn our focus on setups featuring substantial cross sections
for vector-like (s)fermion production at the LHC  connected to
potential novel LHC signatures worthy of investigation. Even though the
LND model has some semblance with the MSSM, it exhibits differences due to the
existence of additional vector-like leptons and down-type quark. Owing to
flavour physics constraints, the vector-like down quark has to be massive and its
coupling to the SM quarks has to be small, which reduces its corresponding LHC
production cross section significantly. Typical LND signals therefore involve
leptonic and often cleaner final states.

We concentrate on vector-like $\tau^{\prime}$ production, in which each extra
lepton dominantly decays into a neutral SM-like Higgs-boson $h$ or $Z$-boson and
a tau lepton,
\be
  p p\to \tau^\prime\ \bar{\tau}^{\prime} \to \tau \bar{\tau} X X
  \qquad\text{with}\qquad X = h, Z \ .
\ee
After accounting for $h$ and $Z$-boson decays, this process could give rise to a
copious production of multileptonic events with small SM backgrounds. We
analyse the four benchmark points defined in Table~\ref{tab:parabm}, that leads
to the production of events containing four first and second generation leptons
($e$ or $\mu$) at the LHC. The tables include the 20 parameters relevant for
collider physics, the most relevant ones being $\varepsilon_{N,3}$, $k_N$,
$\kappa_{L,3}$, $h_N$ and $\varepsilon_E$ as the considered signal involves
vector-like $\tau^\prime$ states.
The corresponding particle spectra are presented in Table~\ref{tab:massp}.

\renewcommand{\arraystretch}{1.2}
\begin{table}
	\begin{center}
		\begin{tabular}{c||c|c|c|c|c|c|c|c}
			 &
			$\mu_L$ [GeV]& $\tan\beta$ & $\varepsilon_{N,3}$  & $k_N$ & $M_0$ [TeV] &  $M_1$ [GeV] & $B_0$ [GeV] & $\mu_D$ [TeV] \\ \hline \hline
			{\bf BP1} &
			$144.9$ & 41.6 & $ -0.045$  & $ 0.013$ & 2.2 & 160.8 & $ 584.64 $ & $ 7.22 $ \\ \hline
			{\bf BP2} &
			 $ 128.9 $ & $ 42.6 $ & $ -0.06 $ & $ -0.18 $ & $ 1.57 $  & $ 168.25 $ & $481.79 $ & $5.47 $\\ \hline
			{\bf BP3} &
			$ 132.42 $ & $ 40.52 $ & $ -0.049$ & $ -0.13 $ & $ 1.65 $  & $ 156.39 $ & $ 452.09 $ & $ 6.03$ \\ \hline
			{\bf BP4} &
			 $ 162.96 $ & $ 25.36 $ & $ -0.035$ & $ 0.0888$ & $ 1.05 $  & $ 206.24 $ & $ 1306.38 $ & $ 4.12 $ \ \\ [.4cm]
 			\multicolumn{7}{c}{}\\
 			 &
 			$M_2$ [TeV] & $\kappa_{L,3}$ [GeV]  & $h_N$ & $\varepsilon_E$ & $A_0$[GeV]& $m^2_{L^\prime,3}$ [TeV$^2$] & $m^2_{D^\prime,3}$ [TeV$^2$] & $m^2_{L,3}$ [TeV$^2$] \\ \hline \hline
 			{\bf BP1} &
 			1.5 & -11.9 &  $ -0.038$  &$ -0.29 $ &  $95.26$ & $ 0.15$ & $ 1.76$ & $ 4.78\!\times\!10^{-3} $ \\ \hline
 			{\bf BP2} &
 			$ 1.05 $ & $ 1.28 $ & $ -6.8 \!\times \!10^{-3} $ & $ -0.16 $ & $ -743.38 $ & $ 2.91\!\times \!10^{-3} $ & $ 1.29 $ & $ 6.86\!\times \!10^{-3}$  \\ \hline
 			{\bf BP3} &
 			$ 1.01 $ & $ 1.16 $ & $ -4.9\!\times \!10^{-3} $ & $ -0.16 $ & $ -516.61 $ & $ 2.09\!\times \!10^{-3} $ & $ 1.55 $ & $ 7.58\!\times \!10^{-3}$  \\ \hline
 			{\bf BP4} &
 			 $ 0.45 $ & $ 15.74 $ & $ -0.106 $ & $ 2.47\!\times \!10^{-3} $ & $ 5.51\!\times \!10^{-4} $ & $ 0.25 $ & $ 23.68 $ & $ 0.20 $  
 			  \\ [.4cm]
 			 \multicolumn{7}{c}{}\\
 			 &
 			 $m^2_{N,3}$ [TeV$^2$] & $m^2_{N^\prime,3}$ [TeV$^2$]& $\mu_N$[GeV] & $m^2_{D,3}$[TeV$^2$] & $\kappa_D $ [GeV] & $ \varepsilon_D$ & $ m^2_{L_5,L,\ 3} $ [TeV$ ^2 $] & $ m^2_{D_5,D,\ 3} $ [TeV$ ^2 $]\\ \hline \hline
 			 {\bf BP1} &
 			 $ 0.32 $ & $ 2.28\!\times \!10^{-2} $& $ 748.84 $ & $ 10.59 $ & $ 0.0 $ & $ -2.38\times10^{-4} $ & $ -9.16\times10^{-5} $ & $ 7.89\times10^{-5} $\\ \hline
 			 {\bf BP2} &
 			 $ 1.32\!\times \!10^{-3} $ & $ 1.11\!\times \!10^{-3} $ & $ 984.28 $ & $ 2.46\!\times \!10^{-3}$ & $ 0.0 $ & $ -9.47\times10^{-4} $ & $ -4.41\times10^{-5} $ & $ -2.91\times10^{-5} $\\ \hline
 			 {\bf BP3} &
 			 $ 1.24\!\times \!10^{-3}$ & $ 8.29\!\times \!10^{-4} $ & $ 975.48$ & $ 3.31\!\times \!10^{-3} $ & $ 0.0 $ & $ -1.08\times10^{-3} $ & $ -4.50\times10^{-5}$ & $ -2.69\times10^{-5} $\\ \hline
 			 {\bf BP4} &
 			 $ 0.29 $ & $ 0.12 $ & $ 1499.71 $ & $ 14.89 $& $ 0.0 $ & $ -3.06\times10^{-4} $ & $ -9.81\times10^{-5} $ & $ 3.86\times10^{-5} $
		\end{tabular}
		\caption{\label{tab:parabm} \it Parameters defining our four representative LND benchmark scenarios {\bf BP1-BP4}.
  The sign of the $\mu$ parameter has been taken positive in all cases.}\end{center}
\end{table}

\begin{table}
	\renewcommand{\arraystretch}{1.2}
	\begin{center}
		\begin{tabular}{c||c|c|c|c|c|c|c}
			 &
			$M_{\tau^\prime}$ [GeV]& $M_{\tilde{\chi}^0_1}$ [GeV] & $M_{\nu^\prime_1}$ [GeV]  & $M_{\nu^\prime_2}$ [GeV] & $M_{b^\prime}$ [TeV] & $M_{\tilde{\chi}^0_2}\cong M_{\tilde{\chi}^\pm_1}$ [TeV] & $ M_{\tilde{e}_1} $ [GeV]  \\ \hline \hline
			{\bf BP1} &
			$ 150.87 $ & $ 157.48 $ & $ 146.71 $ & $ 748.95 $ & $ 7.36 $ & $ 1.53 $ & $ 175.73 $ \\ \hline
			{\bf BP2} &
			$ 133.98$ & $ 164.91 $ & $ 130.22 $ & $ 985.49 $ & $ 5.64 $  & $ 1.09 $ & $ 188.66 $\\ \hline
			{\bf BP3} &
			$ 137.41 $ & $ 153.33 $ & $ 133.68 $ & $ 976.16 $ & $ 6.16 $  & $ 1.05 $ & $ 175.47 $\\ \hline
			{\bf BP4} &
			$ 168.72 $ & $ 202.55 $ & $ 164.21 $ & $ 1500.53$ & $ 4.28 $  & $ 0.48 $ & $ 221.56 $\\
			\multicolumn{7}{c}{}\\
			&
			$M_{\tilde{e}_2}$ [GeV]& $M_{\tilde{\nu}_1}$ [GeV] & $M_{\tilde{\nu}_2}$ [GeV]  & $M_{\tilde{\nu}_3}$ [GeV] & $M_{\tilde{\nu}_4}$ [TeV] & $M_{\tilde{\nu}_5}$ [TeV] & $M_{\tilde{\nu}_6}$ [TeV]  \\ \hline \hline
			{\bf BP1} &
			$ 516.31 $ & $ 169.23 $ & $ 520.67 $ & $ 633.22 $ & $ 1.02 $ & $ 1.47 $ & $ 1.81 $\\ \hline
			{\bf BP2} &
			$ 401.71 $ & $ 193.82$ & $ 401.03 $ & $ 491.87 $ & $ 1.31 $ & $ 1.78 $ & $ 2.39 $\\ \hline
			{\bf BP3} &
			$ 399.22 $ & $ 182.70 $ & $ 396.97 $ & $ 459.45 $ & $ 1.31 $ & $ 1.38 $ & $ 1.99 $\\ \hline
			{\bf BP4} &
			$ 691.04 $ & $ 219.54 $ & $ 691.25 $ & $ 1537.79 $ & $ 1.59 $ & $ 1.80$ & $ 2.39 $\\
		\end{tabular}
		\caption{\label{tab:massp} \it Masses of the particles lighter than $ 2.5 $~TeV for our four representative LND benchmark scenarios {\bf BP1-BP4}.}\end{center}
\end{table}

As soon as all branching ratios are properly included,  final states containing four leptons ($e$ or $\mu$), at least
one hadronic tau and no $b$-tagged jets could yield the largest signal
sensitivity, highlighted in particular by  a low associated background. We make use of
\textsc{MG5\_aMC@NLO} (version 2.6.1)~\cite{Alwall:2014hca} to generate
leading-order hard-scattering events
for both the signal for the four considered benchmarks, and for the
different components of the SM background, for proton-proton collisions at a
centre-of-mass energy of 14~TeV. We generate events for diboson production
(including off-shell effects, once accounting for weak boson leptonic decays),
as well as for the subdominant $t\bar{t}h$, $t\bar{t}Z$ and $t\bar{t}WW$
background contributions. We have additionally verified that triboson
background contributions were negligible. In our simulations, we rely on the
four-flavour number scheme, making use of the leading-order set of NNPDF2.3 parton
densities~\cite{Ball:2012cx}. We include taus when enforcing weak boson leptonic
decays, and allow for the presence of up to two extra partons in the final
state. The multipartonic contributions are then merged following the MLM
prescription~\cite{Mangano:2006rw}. Parton showering and hadronisation are
performed within the \textsc{Pythia8} (version 8.230)
framework~\cite{Sjostrand:2007gs} and we simulate the response of the detector
by means of \textsc{Delphes 3} (version 3.4.1)~\cite{deFavereau:2013fsa}. We
modify slightly  the default CMS detector parameterisation that relies
on the \textsc{FastJet} program (version 3.2.1)~\cite{Cacciari:2011ma} for jet
reconstruction on the basis of the anti-$k_T$ algorithm~\cite{Cacciari:2008gp}, 
with a radius parameter set to $R=0.5$. Our modifications imply a tau-tagging
efficiency fixed to 60\%, for a mistagging rate of a light-jet as a hadronic
tau set to 1\% (this configuration matches average performances after the
object selection enforced below). In contrast, we consider standard $b$-tagging
performance as implemented in the default CMS parameterisation~\cite{%
Chatrchyan:2012jua}.

\begin{figure}
  \centering
   \includegraphics[width=0.32\columnwidth]{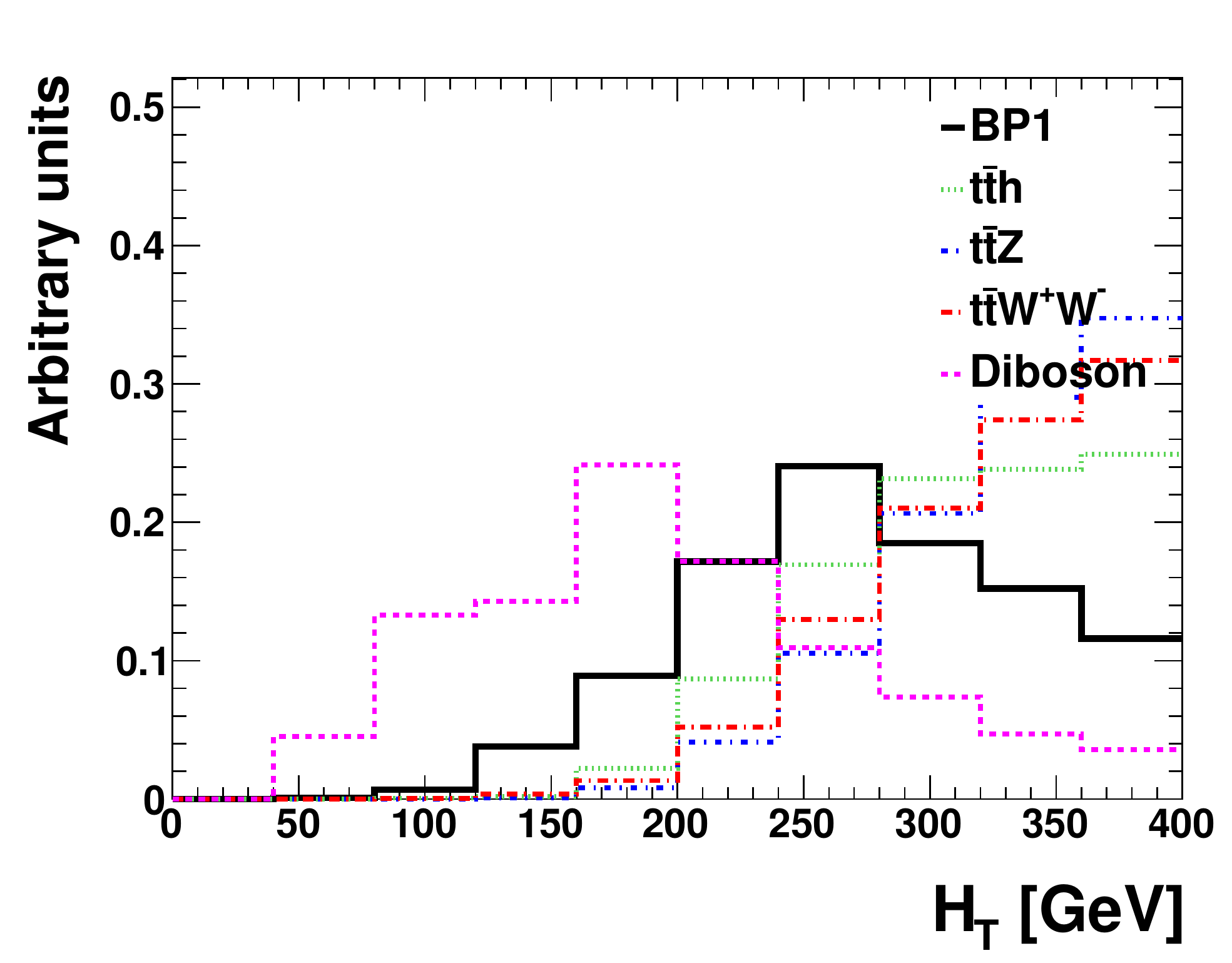}
   \includegraphics[width=0.32\columnwidth]{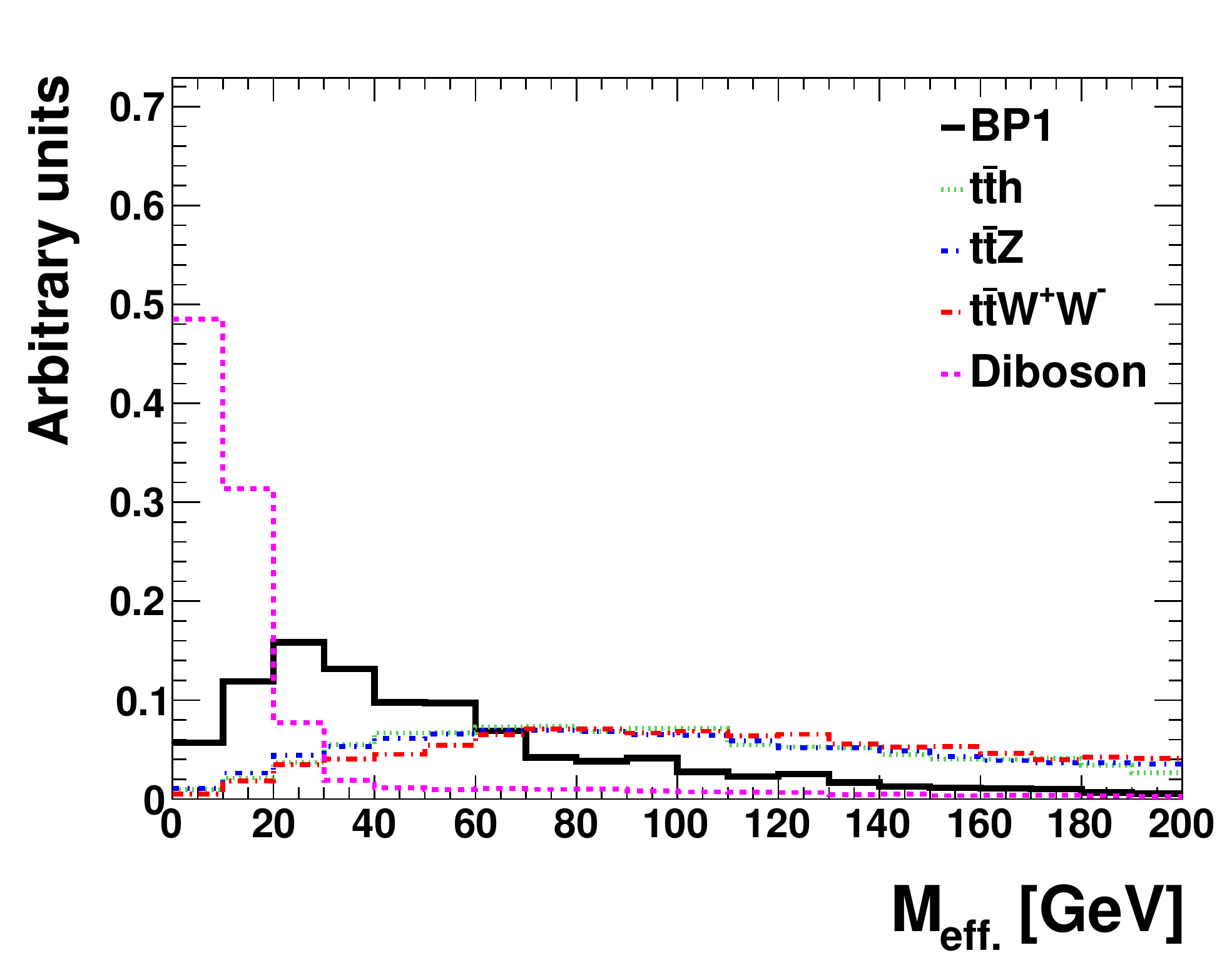}
   \includegraphics[width=0.32\columnwidth]{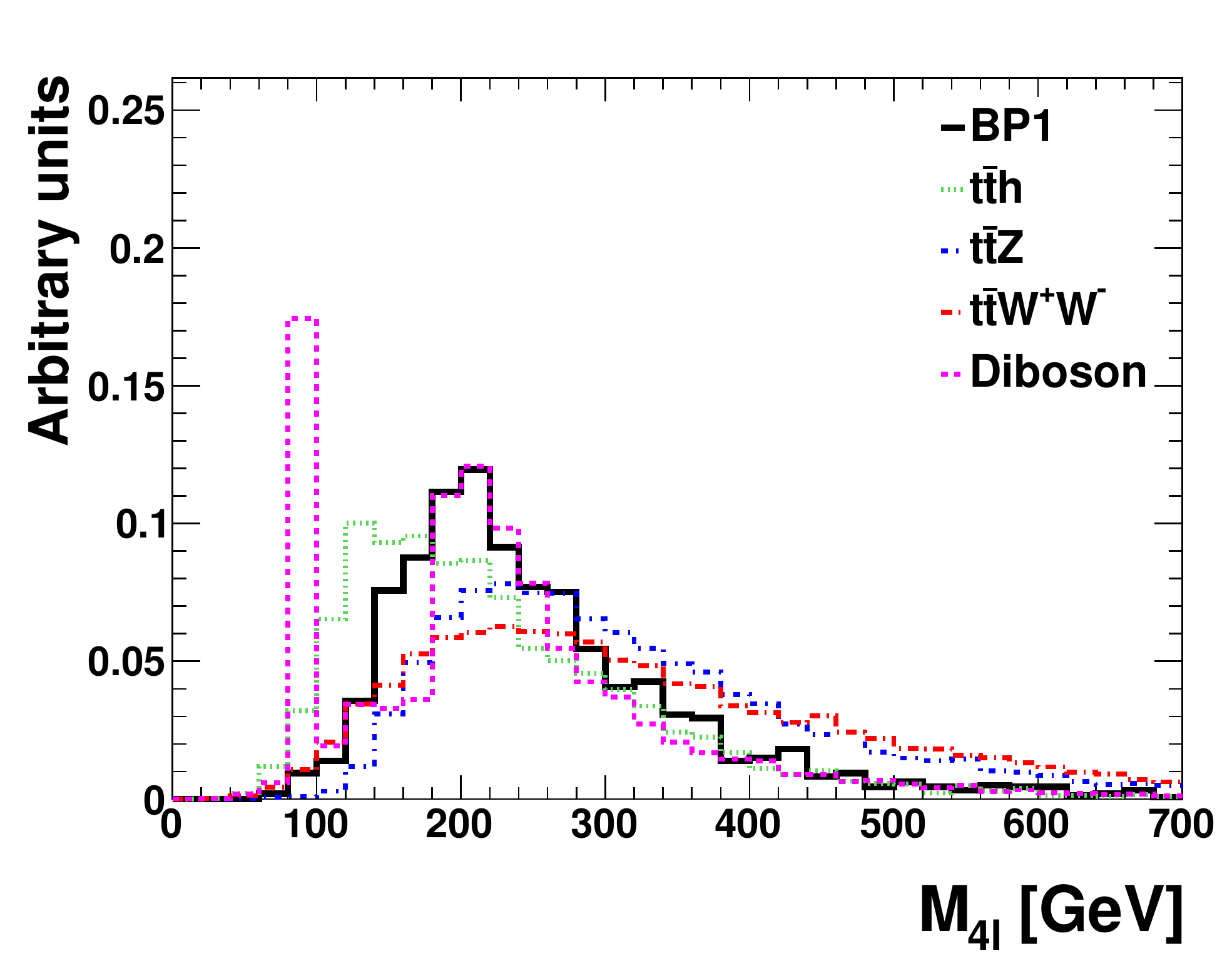}
   \caption{\it Distributions in the $H_T$, $M_{\rm eff}$ and $M_{4\ell}$
     observables for the {\bf BP1} benchmark scenario and the dominant
     contributions to the SM background, once the preselection cuts of
     Eq.~\eqref{eq:presel2} have been imposed. The normalisation is arbitrary.}
       \label{fig:vars}
\end{figure}

We define the relevant reconstructed object candidates by imposing transverse
momentum ($p_T$) and pseudorapidity ($\eta$) conditions on the leptons
($\ell=e$, $\mu$), hadronic taus ($\tau_h$) and light and $b$-tagged jets ($j$
and $b$),
\be
  p_T^{\ell}  \geq 10~{\rm GeV}\ \text{and}\ |\eta^{\ell}| < 2.5 \ ,\qquad
  p_T^{\tau_h}\geq 20~{\rm GeV}\ \text{and}\ |\eta^{\tau}| < 2.5 \ ,\qquad
  p_T^{j,b}   \geq 30~{\rm GeV}, |\eta^j| < 4.5\ \text{and}\
   |\eta^b| < 2.5\ .
\label{eq:presel}\ee
We moreover require lepton isolation by imposing that the total hadronic
activity within a cone of radius $\Delta R=0.5$ around any electron (muon) is
smaller than 12\% (25\%) of the lepton $p_T$, and that all reconstructed leptons
are separated from each other by an angular distance, in the transverse plane,
of at least $R=0.5$. We then preselect events by
constraining the number of reconstructed final-state electrons and muons
($N^\ell$), hadronic taus ($N^{\tau_h}$) and $b$-tagged jets ($N^b$), to be
\be
  N^{\ell} \ge 4\ , \qquad N^b = 0 \qquad\text{and}\qquad N^{\tau_h} \ge 1\ .
\label{eq:presel2}\ee
We then investigated a large set of observables and found that the
most discriminatory ones are the total transverse activity $H_T$ (the
scalar sum of the $p_T$ of all reconstructed visible objects), a modified version
of the effective mass $M_{\rm eff}$ (the scalar sum of the $p_T$ of all jets and
the missing transverse energy $\slashed{E}_T$), and the invariant mass 
$M_{4\ell}$ of the system made of the four hardest leptons. The distributions in
these variables are shown in Figure~\ref{fig:vars} for both the different
background contributions and the illustrative {\bf BP1} benchmark scenario.
Upon scrutinising these variables, we select events for which
\be
  H_T > 250~{\rm GeV}, \qquad M_{\rm eff} > 30~{\rm GeV} \qquad\text{and}
    \qquad M_{4\ell} > 200~{\rm GeV}.
\ee
Whilst further optimisation is possible, these choices allow for a good enough
background rejection in the context of the four considered benchmark selections,
as illustrated by the detailed cutflow charts shown in Table~\ref{tab:cutflow}
for 3~ab$^{-1}$ of LHC collisions at 14~TeV.

\begin{table}
 \renewcommand{\arraystretch}{1.4}
 \setlength\tabcolsep{7pt}
 \begin{center}
  \begin{tabular}{c l|c|cccc}
   Step&Requirements & Background & {\bf BP1} &  {\bf BP2} & {\bf BP3} &
     {\bf BP4}\\\hline
   0 & Initial      &  $5.3\times10^9$ & $1.4\times10^8$ & $2.0\times10^8$ &
     $2.1\times10^8$ & $1.7\times10^8$\\
   1 & Preselection          & 149 & 241 & 336 & 336 & 282\\
   2 & $H_T > 250$ GeV       & 101 & 183 & 240 & 247 & 236\\
   3 & $M_{\rm eff}>30$ GeV  &  39 & 117 & 120 & 125 & 156\\
   4 & $M_{4\ell} > 200$ GeV &  34 &  90 &  95 & 100 & 118\\\hline
   & \multirow{2}{*}{3 ab$^{-1}$ (300 fb$^{-1}$)}& $s$
       & $10.06\sigma$ ($4.59\sigma$) & $10.57\sigma$ ($4.82\sigma$)
       & $11.13\sigma$ ($5.08\sigma$) & $13.16\sigma$ ($6.00\sigma$)\\
   && $Z_A$
       & $6.70\sigma$ ($3.39\sigma$) & $6.94\sigma$ ($3.53\sigma$)
       & $7.22\sigma$ ($3.68\sigma$) & $8.14\sigma$ ($4.19\sigma$)\
  \end{tabular}
  \caption{\it Impact of our event selection strategy on the SM background and the
    four considered benchmark scenarios. For each cut, we provide the expected
    number of surviving events for $\mathcal{L}=3$~ab$^{-1}$ of LHC collisions
    at a centre-of-mass energy of 14~TeV. We also quote the corresponding
    significances $s$ and $Z_A$ defined in Eq.~\eqref{eq:sigs}, including a 20\%
    systematic uncertainty on the background. We additionally indicate, in
    parentheses, the significances for a lower luminosity of 300~fb$^{-1}$.}
  \label{tab:cutflow}
 \end{center}
\end{table}

We evaluate the sensitivity of the high-luminosity LHC to our different
benchmark scenarios by computing two significance indicators $s$ and $Z_A$
defined by~\cite{Cowan:2010js}
\be
  s = \frac{S}{\sqrt{B+\sigma_B^2}}\qquad\text{and}\qquad
  Z_A= \sqrt{ 2\left(
     (S+B)\ln\left[\frac{(S+B)(S+\sigma^2_B)}{B^2+(S+B)\sigma^2_B}\right] -
     \frac{B^2}{\sigma^2_B}\ln\left[1+\frac{\sigma^2_BS}{B(B+\sigma^2_B)}\right]
     \right)} \ ,
\label{eq:sigs}\ee
with $S$ and $B$ being respectively the total number of signal and background
events surviving the selection. Our results assume a 20\% systematic uncertainty
on the SM background, $\sigma_B = 0.2 \times B$.  In
Table~\ref{tab:cutflow} we compare the expectation for both 300~fb$^{-1}$ and 3~ab$^{-1}$
of LHC collisions. The discovery prospects are in all cases very promising, so
that the presence of the vector-like leptons offers good handles on LND models.
It will, however, be challenging to conclude about the realisation of the model in
nature without getting a grip on the supersymmetric part of the spectrum with
the LHC alone.

This could for instance be achieved by investigating the impact of the
searches for supersymmetry through its monojet and multijet plus missing energy
signatures~\cite{Aaboud:2016tnv,Aaboud:2016zdn,Aaboud:2017phn,Aaboud:2017vwy}.
In all cosmologically-favoured LND setups analysed in Section~\ref{sec:DM}, such
signals arise from the pair production of squarks and/or gluinos. However, the
lightest squarks have generally masses of about 2~TeV or more, so that the
corresponding cross sections are negligibly small, especially after imposing the
presence of at least one very hard jet in the final state. Exceptional scenarios
nevertheless exist, in which lighter coloured superpartners are featured. In
this case, the relevant cross sections are of ${\cal O}(1)$~fb, which is too
small to yield any hope of observing a hint for the signal. We have, in addition,
evaluated the cross sections associated with electroweakino pair production. For
all scenarios favoured by cosmological data, they reach at most 0.1~fb, when
the branching-ratio-favoured hadronic final states are considered. The direct
observation of a supersymmetric signal at the LHC will, hence, be very unlikely. 
Thankfully, as demonstrated in this work, the presence of vector-like leptons in the LND model provides additional observational handles 
which are complementary to cosmological and astrophysical probes.

Our collider analysis focused on decays into vector-like $\tau^\prime$ leptons,
yielding signals with four leptons (electrons or muons), one hadronic tau and no
$b$-jets. This constitutes a rather unique signature that is neither probed by
multilepton analyses, which target final states with three or more leptons but no
taus~\cite{Sirunyan:2017qkz}, nor by conventional searches for vector-like leptons,
which include either three or more light leptons and no taus, or two light
leptons and a single tau, as in Ref.~\cite{CMS:2018cgi}. Perhaps the closest
experimental study to our proposal is the ATLAS analysis of
Ref.~\cite{Aaboud:2018zeb}. However, the information provided in their Tables 4
and 6 indicate that not a single one of their signal region matches ours.
Hopefully, further analyses at the LHC will allow for the
investigation of a wider range of parameter space and probe additional
multilepton signals.

\section{Summary and conclusions}
\label{sec:conclusions}
In this work we presented an analysis of an extension of the MSSM along lines
suggested by supersymmetric grand unification. The minimal superfield content of
the MSSM is enlarged by the addition of a complete $\bf{5}\oplus {\bf{\overline{5}}}$
representation of $SU(5)$, which leads to vector-like pairs of down-type
quark and lepton supermultiplets after the breaking of the grand-unified
symmetry. We moreover include a vector-like pair of
singlet neutrino supermultiplets. The model then provides, in addition to the
lightest neutralino, a new potential candidate for dark matter compared with
other GUT-inspired supersymmetric models with vector-like fermions, namely the scalar
superpartner of the singlet vector-like neutrino (the latter becomes a sterile neutrino).

We set up the model allowing mixing of the new vector-like states with the third generation of down-type quarks and leptons only (to avoid unwanted flavour-changing effects), and investigated its consequences by scanning over the free parameters over a wide range. The parameter space is then restricted by imposing collider mass limits,
the compatibility with a complete set of Higgs-sector-related measurements, and
constraints originating from electroweak precision tests, lepton flavour
violation and $B$-physics. We first investigated the dark matter candidates
featured by the model and imposed restrictions from the requirement to reproduce
the observed dark matter abundance in the Universe and the absence of a signal
in direct and indirect dark matter detection experiments.

The neutralino DM candidates turn out to have fairly similar properties as in
the MSSM. In particular, in this model  the bino-(vector-like) electron-selectron coupling is a gauge coupling, and thus proportional to the hypercharge $Y$. The bino-like neutralino annihilation cross-section into vector-like fermions is, then, proportional to $Y^4$, and since the vector-like (s)fermions that are charged under $U(1)_Y$ belong to $SU(2)_L$ doublets they carry a hypercharge of $1/2$. This, in turn, implies that the overall bino annihilation cross-section is suppressed in comparison to models containing vector-like $SU(2)_L$ singlets. As we showed explicitly, this suppression hinders these novel (with respect to the MSSM) annihilation channels from providing an efficient mechanism for dark matter depletion in the early Universe and deprives heavier binos of the possibility to be viable DM candidates.

While sneutrinos can in principle be an arbitrary admixture of the MSSM ($SU(2)_L$ doublet) left-handed sneutrinos and the vector-like left- and right-handed $SU(2)_L$ doublet or singlet ones, and while the mostly doublet-like sneutrino dark matter option is entirely compatible with the requirement to reproduce the observed relic density,  direct detection experiments exclude such scenarios. This means that only singlet-like sneutrinos survive, making this scenario difficult to differentiate from other models where the right-handed sneutrino is included on an {\it ad-hoc} basis, or is required by the symmetry of the model. However, as we argued, given all current experimental constraints sneutrino dark matter in this scenario can further find motivation from naturalness arguments, which constitutes an interesting twist of the LND model with respect to other minimal GUT-motivated MSSM extensions with vector-like fermions,  like the QUE and QDEE models. 

Lastly, the model shows some interesting prospects in collider signals. We devised
benchmark points with substantial cross sections to vector-like final states, to
unravel the corresponding signals that are absent from the MSSM. We concentrated on
vector-like $\tau^{\prime}$ production, decaying dominantly into a neutral boson
(either the SM-like Higgs boson $h$ or the $Z$ boson) and a tau lepton. We have
demonstrated that the pair-production of a pair of vector-like $\tau^{\prime}$
yields a multilepton signature that could distinguish this model in the future high-luminosity runs of the LHC,
for both luminosities of 300~fb$^{-1}$ and 3~ab$^{-1}$. Taken together with the
cosmological implications of the model, this signal could provide a way to assess its viability.

\begin{acknowledgments}
We thank F.~Staub and A.~Pukhov for their respective help with the {\sc Sarah}
and {\sc MicrOMEGAs} packages, as well as S.~Jain for helpful discussions on
the collider front. This research was enabled in part by support provided by
the High Performance Computing (HPC) server Guillimin in McGill
University (\href{www.hpc.mcgill.ca}{hpc.mcgill.ca}) and Compute Canada
(\href{www.computecanada.ca}{computecanada.ca}). JYA and MF acknowledges NSERC
for partial financial support under grant number SAP105354 and JYA acknowledges the 
hospitality of the LPTHE, Sorbonne University and INFN -- Laboratori Nazionali di Frascati where part of this
work was completed. The work of BF and AG is partly supported by French state funds managed by the Agency Nationale de
la Recherche (ANR), in the context of the LABEX ILP (ANR-11-INDEX-0004-02,
ANR-10-LAB-63). SB is supported by a Durham Junior Research Fellowship 
COFUNDed by Durham University and the European Union, under grant agreement number 609412,
and acknowledges the hospitality of the CERN Theory Department where part of this
work was completed.
\end{acknowledgments}

\providecommand{\href}[2]{#2}\begingroup\raggedright\endgroup

\end{document}